\documentclass[prd,preprint,showpacs,preprintnumbers,nofootinbib,superscriptaddress]{revtex4}

\usepackage{amsmath}
\usepackage{amssymb}
\usepackage{graphicx}
\usepackage{axodraw}
\usepackage{enumerate}
\newcommand{\del}{\partial} 
\newcommand{\Tr}{{\rm Tr}} 

\newcommand{\sla}[1]{{/\hskip-0.55em{}#1}{}}
\newcommand{\Sla}[1]{{\hskip+0.2em/\hskip-0.75em{}#1}{}}

\begin{document}
%----------------------------------------------------------------
\date{\today}

\preprint{RIKEN-TH-61}
\preprint{hep-ph/0512288}

%================================================================

\title{%
Automated Calculation Scheme for $\alpha^n$ Contributions of QED 
to Lepton $g\!-\!2$: Generating Renormalized Amplitudes for Diagrams 
without Lepton Loops}

%----------------------------------------------------------------

\author{T.~Aoyama}
\affiliation{Theoretical Physics Laboratory, RIKEN, Wako, Saitama, Japan 351-0198 }

\author{M.~Hayakawa}
\affiliation{Theoretical Physics Laboratory, RIKEN, Wako, Saitama, Japan 351-0198 }

\author{T.~Kinoshita}
\affiliation{Laboratory for Elementary Particle Physics, Cornell University, Ithaca, New York 14853, U.S.A. }

\author{M.~Nio}
\affiliation{Theoretical Physics Laboratory, RIKEN, Wako, Saitama, Japan 351-0198 }

%================================================================
%================================================================
\begin{abstract} 
Among 12672 Feynman diagrams contributing to 
the electron anomalous magnetic moment at the tenth order, 
6354 are the diagrams 
having no lepton loops, \textit{i.e.}, those of quenched type. 
Because the renormalization structure of these diagrams 
is very complicated, 
some automation scheme is inevitable to calculate them. 
We developed an algorithm to write down 
FORTRAN programs for numerical evaluation of these diagrams, 
where the necessary counterterms to subtract out 
ultraviolet subdivergence are generated according to 
Zimmermann's forest formula. 
Thus far we have evaluated crudely integrals of 2232 
tenth-order vertex diagrams which require vertex renormalization 
only.
Remaining 4122 diagrams, which have ultraviolet-divergent 
self-energy subdiagrams and infrared-divergent subdiagrams, 
are being evaluated by giving small mass $\lambda$ to photons 
to control the infrared problem.
\end{abstract}
%================================================================

%================================================================

%----------------------------------------------------------------

% 13.40.Em Electric and magnetic moments
% 14.60.Cd Electrons (including positrons)
% 14.70.Bh Photons
% 11.15.Bt General properties of perturbation theory
% ( 12.20.-m Quantum electrodynamics )
% 12.20.Ds Specific calculations

%----------------------------------------------------------------

\maketitle
%\tableofcontents

%================================================================

%================================================================
\section{Introduction} 
%================================================================
\label{sec:introduction}

The anomalous magnetic moment of the electron, 
also called the electron $g\!-\!2$, 
is one of the most fundamental quantities of particle physics. 
Since its discovery in 1947 \cite{kusch}
it has been measured with steadily increasing precision 
\cite{rich,VanDyck:1987ay}. 
The best values of $g\!-\!2$ of the electron and the positron 
available in the literature \cite{VanDyck:1987ay} 
\begin{equation}
\begin{aligned}
	a_{e^-}({\rm exp}) 
	&=
	1\ 159\ 652\ 188.4\ (4.3) \times 10^{-12}
	\,, \\
	a_{e^+}({\rm exp}) 
	&= 
	1\ 159\ 652\ 187.9\ (4.3) \times 10^{-12}
\label{eq:expValue} 
\end{aligned}
\end{equation}
were obtained by the Penning trap experiment. 
Here $a_e \equiv \frac{1}{2} (g-2)$ 
and the numerals in each parenthesis represent uncertainty 
in the last few digits of the respective values. 
The consistency of $a_{e^-}$ with $a_{e^+}$ in (\ref{eq:expValue}) 
within the experimental accuracy exhibits 
that CPT is a very good symmetry of the universe. 

At present a new experiment is being carried out by a Harvard group 
using a new trap with cylindrical cavity 
\cite{Gabrielse},
which is capable of controlling the electron--cavity-wall resonance 
with the help of analytical calculation \cite{brown}.
This experiment will reduce the measurement uncertainty of 
(\ref{eq:expValue}) substantially. 
It will enable us to test the validity of QED to a very high degree 
and to determine the fine structure constant $\alpha$ 
to an unprecedented precision of $7 \times 10^{-10}$ or better, 
which is an order of magnitude better 
than the best non-QED value available at present \cite{wicht}. 

Of course, such a feat requires availability of theoretical calculation 
of matching precision. 
Within the framework of the Standard Model the QCD and weak interaction 
parts of the corrections to $a_e$ are known to be so small that 
their uncertainties do not affect the determination of $\alpha$ 
even with the expected precision of the new Harvard experiment. 
The uncertainties due to the QED contributions induced by the virtual 
propagation of muon and tau-lepton beginning at the fourth-order 
($\alpha^2$) are also known to be negligible within the current precision. 
Thus the electron $g\!-\!2$ within the experimental precision of our 
current interest is determined almost entirely by the electron-photon 
interaction and can be regarded as a function of $\alpha$ alone. 

The latest evaluation of $a_e$ in the Standard Model, including 
the hadronic vacuum polarization contribution, 
hadronic light-by-light-scattering contribution, 
the electroweak effect, 
and small QED contribution from virtual muon and tau-lepton loops 
is \cite{kn2} 
\begin{equation} 
	a_e = 
	1\ 159\ 652\ 175.86\ (0.10)\ (0.26)\ (8.48) 
	\times 10^{-12}
	\,,
\label{eq:a_e_KN}
\end{equation} 
where the uncertainties stem from 
(i) the remaining numerical uncertainty of the $\alpha^4$-term 
\cite{kn2}, 
(ii) the crudely estimated uncertainty of the $\alpha^5$-term 
\cite{CODATA:2002}, 
and (iii) that of the best non-QED $\alpha$ available at present, 
which is measured by the atom interferometry \cite{wicht} 
combined with the cesium $D_1$ line measurement by the 
frequency comb technique \cite{udem}, 
\begin{equation} 
	\alpha^{-1}(h /M_{{\rm Cs}}) 
	= 
	137.036\ 000\ 3\ (10)  
	\quad [7.4\ {\rm ppb}]
	\,.
\label{eq:alpha_cs}
\end{equation} 

An important byproduct of the study of the electron $g\!-\!2$ is 
that a more precise $\alpha$ can be obtained by combining 
the measurement (\ref{eq:expValue}) and the theory of $a_e$, 
which yields \cite{kn2} 
\begin{equation}
	\alpha^{-1}(a_e) =
	137.035\ 998\ 834\ (12)\ (31)\ (502)
	\quad [3.7\ {\rm ppb}]
	\,, 
\label{eq:alpha_ae_seattle}
\end{equation}
where the uncertainties $12$ and $31$ are due to the $\alpha^4$ and 
$\alpha^5$ terms, and $502$ comes from the experiment (\ref{eq:expValue}). 
The new Harvard experiment of $a_e$ is expected to reach a precision 
an order of magnitude better than that of (\ref{eq:expValue}). 
The $\alpha^5$ term will then become the largest source of unresolved 
systematic errors for obtaining $\alpha$ from $a_e$. 
Thus an explicit evaluation of the $\alpha^5$ term is urgently 
needed for further improvement of $\alpha(a_e)$. 

The pure QED contribution can be written as 
\begin{equation}
	a_e({\rm QED})
	=A_1^{(2)}\left(\frac{\alpha}{\pi}\right)
	+A_1^{(4)}\left(\frac{\alpha}{\pi}\right)^2
	+A_1^{(6)}\left(\frac{\alpha}{\pi}\right)^3
	+A_1^{(8)}\left(\frac{\alpha}{\pi}\right)^4
	+\cdots
	\,.
\end{equation}
The coefficients are evaluated by the perturbation theory. 
By now the first four of them have been obtained \cite{kn2}: 
\begin{equation}
\begin{aligned}
	A_1^{(2)\phantom{0}} &=
	\phantom{-}0.5 \,, \\
	A_1^{(4)\phantom{0}} &=
	-0.328\ 478\ 965 \cdots \,, \\
	A_1^{(6)\phantom{0}} &=
	\phantom{-}1.181\ 241\ 456 \cdots \,, \\
	A_1^{(8)\phantom{0}} &=
	-1.728\ 3\ (35) \,.
\end{aligned}
\label{eq:coefficients_a(QED)}
\end{equation}
$A_1^{(10)}$ has not yet been evaluated. 
An educated guess is that it may be found within the range 
$(-3.8, 3.8)$ \cite{CODATA:2002}. 

The first theoretical calculation of $a_e$ was carried out analytically 
by Schwinger in 1948 \cite{schwinger}. 
The number of Feynman diagrams involved was just one in this case. 
The excellent agreement of his calculation with the measurement \cite{kusch} 
was one of the pivotal triumphs of renormalization theory of QED, 
which was just being developed. 
Refinement of theory to the fourth-order involves seven Feynman diagrams. 
It took more than 7 years before the analytic value of $A_1^{(4)}$ 
was obtained in 1957 \cite{som-pet}. 
Analytic evaluation of $A_1^{(6)}$ is far more challenging 
requiring evaluation of 72 Feynman diagrams. 
It took effort of many physicists and many years of hard work and 
was completed only in 1996 \cite{lap-rem}.

The numerical evaluation scheme was developed by one of the authors 
(T.~K.) and Cvitanovi{\'c} for the evaluation of the sixth-order 
contribution 
\cite{Cvitanovic:1974uf,Cvitanovic:1974sv,Cvitanovic:1974um} 
and extended later to the eighth-order \cite{kl1}. 
Up to now $A_1^{(8)}$ was calculated only numerically, which involves 
the evaluation of 891 Feynman diagrams. 
Although the initial crude result was published in 1974 \cite{kl1}, 
improvement of numerical precision required many years of extensive 
computation and the final result was published only recently \cite{kn2}. 
From the viewpoint of obtaining $A_1^{(10)}$ 
the numerical integration approach is the only practical choice 
at present.

The contribution to the $\alpha^5$ term of the electron $g\!-\!2$ 
comes from $12672$ vertex-type Feynman diagrams, 
which can be categorized into 6 sets according to their structures 
and classified further into 32 gauge-invariant subsets. 
See Appendix~\ref{sec:classification} and Ref.~\cite{kn3} 
for the details of classification. 
Most subsets that contain closed lepton loops have relatively 
simple structure and are calculable by a slight extension of 
the method developed in 
Refs.~\cite{Cvitanovic:1974um} and \cite{kl1}. 
As far as the muon $g\!-\!2$ is concerned, they cover the subsets 
that give rise to the leading contributions. 
Thus far 17 of 32 subsets have been evaluated and reported 
in Refs.~\cite{kino1} and \cite{kn-radcor05}. 
Detail of these works is presented in Ref.~\cite{kn3}. 

For the electron $g\!-\!2$, however, none of 32 subsets is dominant 
so that all must be evaluated. 
A particularly difficult one is Set V, a huge set consisting of 6354 
vertex diagrams, all of which have pure radiative corrections and 
no closed lepton loops 
%(see Fig.~\ref{fig:setv}).
(see Fig.~\ref{fig:set5} in Appendix~\ref{sec:classification}). 
Throughout this paper these diagrams will be referred to 
as ``quenched-type (q-type)'' since they are analogous to the 
so-called quenched diagrams of QCD. 
The difficulty of Set V stems from the fact that many of them have 
very large number of ultraviolet (UV) and infrared (IR) divergences. 
This makes the previous approach highly impractical 
since it runs into an extremely sever logistic problem. 
Unless this is solved, it will be close to impossible for mortals 
to deal with Set V and hence the entire tenth-order electron $g\!-\!2$ 
without making mistake. 

The purpose of this paper is to present our solution to this 
very difficult problem. 
We have developed a scheme of automatic code generation 
which enables us to generate renormalized integrals for 
all diagrams of Set V with a breathtaking speed. 
Outputs of this code are ready to be integrated by numerical means.

We begin by exploiting the equation derived from the Ward-Takahashi 
identity, which relates the sum of a set of vertex diagrams to 
a self-energy-like diagram \cite{Cvitanovic:1974um}. 
This relation together with the time-reversal invariance of QED 
enables us to reduce the number of independent integrals of 
Set V diagrams drastically from 6354 to 389. 
The method starting from the Ward-Takahashi (W-T) summed diagrams 
is called \textit{Version~A}, while the conventional approach by 
vertex diagrams is referred to as \textit{Version~B} 
for the sake of distinction \cite{kn1}. 

The systematic scheme for constructing numerical integration code 
consists of the following steps.
\begin{enumerate}[(I)]
\item 
Identify the diagrams contributing to the electron $g\!-\!2$ 
and their UV- and IR-divergent subdiagrams. 

\item 
Carry out momentum space integration 
exactly using a home-made integration table 
and convert it into an integral 
over Feynman parameters. 
The result is expressed symbolically as a function of quantities 
$U$, $B_{ij}$, and $A_j$, 
which are homogeneous polynomials of Feynman parameters. 
We call them \textit{building blocks}. 

\item 
Find the explicit forms of $U$ and $B_{ij}$ which are determined from 
the \textit{topological structure} of the Feynman diagram ${\cal G}$ 
obtained by removing all external lines and 
disregarding the distinction between an electron line and a photon line. 

\item 
To prepare for numerical integration on computer 
UV and IR divergences must be removed from the integrand beforehand. 
In Ref.~\cite{Cvitanovic:1974um} 
a regularization scheme was developed in which divergences 
are eliminated by point-by-point subtraction 
by counterterms which are derived from the original integrand 
by simple power-counting rules. 
This scheme is denoted as 
\textit{K}-operation for the UV divergence and 
\textit{I}-operation for the IR divergence 
\cite{Cvitanovic:1974sv,Kinoshita_book}. 

\item 
Counterterms thus constructed can be identified with only 
the UV divergent parts of the renormalization constants 
so that the result of step (IV) is not fully equivalent 
to the standard on-shell renormalization. 
The difference between full renormalization and intermediate 
renormalization 
must therefore be evaluated by summing up all subtraction terms. 

\end{enumerate}

The scheme (I)--(V) itself is completely general and applicable 
to any order of perturbation. 
Step (II) was quite difficult already in the sixth-order 
and still harder in the eighth-order. 
It was carried out entirely by computer with the help of 
algebraic manipulation programs such as 
SCHOONSCHIP \cite{veltman} and FORM \cite{vermaseren}. 
Steps (I), (III), (IV) and (V) were simple enough in the sixth-order 
case and still manageable in the eighth-order case 
to be executed by hand manipulation. 

It is evident, however, that such a pedestrian approach 
is no longer adequate for the calculation of the tenth-order 
diagrams so that a highly automated approach is required. 
This applies not only to the step (II) but to all other steps. 
It turns out that the automation scheme can be formulated quite 
efficiently for the q-type diagrams by making full use of their 
inherent properties. 
For Step (I) a systematic procedure for the generation of diagrams 
and the identification of UV divergent subdiagrams is possible. 
The graph-theoretical notions are easily identified for 
this type of diagrams, which enables the automated construction 
of topological quantities in Step (III). 

The UV subtractions in Step (IV) can be organized by following 
the Zimmermann's \textit{forests} of subdiagrams exactly 
\cite{Zimmermann:1969jj}. 
The forests are constructed as combinations of subdiagrams 
identified in Step (I). 

This enables us to write a code which controls all steps 
(I), (II), (III), and (IV) automatically. 
Namely, we have obtained a code which turns an input of 
\textit{single-line} information characterizing the structure 
of a Feynman diagram into a fully renormalized 
Feynman parametric integral. 

Thus far we have obtained FORTRAN codes of renormalized integrals 
for 2232 vertex diagrams which contain vertex renormalization 
subdiagrams only. 
Crude evaluation by the Monte-Carlo integration routine 
VEGAS \cite{lepage} shows that our scheme works well as expected. 

The next step is to evaluate the remaining 4122 diagrams 
which have self-energy subdiagrams. 
These diagrams have also IR divergences. 
The simplest way to deal with the IR problem is to give 
a small mass $\lambda$ to photons, which can be implemented 
with a minor extension of the automating code. 
Of course, the numerical result will have an uncertainty of 
order $\lambda$. 
It may also suffer from non-negligible digit-deficiency problem 
commonly encountered in numerical integration \cite{kn1}. 
Nevertheless it will be good enough for getting crude values of 
$A_1^{(10)}$ so that we follow this approach as the first step. 

This scheme has thus far been tested successfully with the 
sixth-order q-type diagrams and reproduced the analytic 
result after proper treatment of residual renormalization terms. 
The next step is to check the eighth-order q-type diagrams, 
which is numerically known. 
It seems successful so far except for a few diagrams which 
have severe IR divergences. 
Those diagrams may require minor modifications to the present 
automation code. 
After these exercises we will tackle the tenth-order problem. 

To obtain a result independent of $\lambda$ it is necessary to extend 
the automating code to include IR subtraction terms constructed 
in a manner similar to that of UV counterterms. 
To complete this calculation we must evaluate the contribution of 
residual renormalization terms, which consists of 
integrals of up to eighth-order for 6804 UV-divergent subdiagrams. 
The result of these works will be reported in the subsequent papers.

%----------------------------------------------------------------
This paper is organized as follows. 
In Section~\ref{sec:general} 
we briefly review the parametric integral formalism 
to obtain the anomalous magnetic moment of leptons 
in \textit{Version A} approach. 
In Section~\ref{sec:UVsubtraction} 
we describe \textit{K}-operation for the subtraction of 
UV divergence which derives from a single UV-divergent subdiagram. 
We then proceed to the case involving more than one subdiagram 
in Section~\ref{sec:forest_formula} 
in relation to the \textit{forest} structures. 
It is one of the essential ingredients for our automated scheme. 
In Section~\ref{sec:q-type} and \ref{sec:q-type_UV} 
we focus on a specific type of diagrams, namely, q-type diagrams. 
We discuss their intrinsic properties 
in Section~\ref{sec:q-type}. 
We see 
in Section~\ref{sec:q-type_UV} 
that use of these properties allows us to develop algorithms 
to identify UV divergences in terms of all relevant forests. 
It enables us to construct the renormalized amplitude 
in an automated manner. 
In Section~\ref{sec:flow} 
we show the whole flow of our automated scheme in detail. 
Section~\ref{sec:conclusion} 
is devoted to the conclusion and discussions. 
In Appendix~\ref{sec:classification} 
we show the classification of diagrams that contribute to 
$A_1^{(10)}$. 
Appendix~\ref{sec:detail} contains useful formulae 
for computing the basic building blocks 
of the Feynman integrals which can be readily adapted to 
the programming languages such as C++ and FORTRAN. 

%================================================================

%================================================================
\section{General Formalism}
%================================================================
\label{sec:general}

In this section we present the general formalism for evaluating 
QED contribution of anomalous magnetic moment of lepton 
in perturbation theory. 
It is a brief summary of the literature
\cite{Cvitanovic:1974um,Kinoshita_book}, 
which is included here to provide concrete prescription of formulation 
for our automation scheme presented in the latter part of this paper, 
and also to make this article self-contained.

%----------------------------------------------------------------
\subsection{Anomalous magnetic moment of lepton}
\label{sec:general:ae}

The magnetic property of a lepton can be studied through 
examining its scattering by a static magnetic field. 
The amplitude of this process including interactions with 
the virtual photon fields can be represented as follows, 
by taking account of the gauge symmetry, invariances 
under Lorentz, C, P, and T transformations: 
\begin{equation}
	e\bar{u}(p^{\prime\prime})\left[
	\gamma^\mu\,F_1(q^2) 
	+ \frac{i}{2m}\sigma^{\mu\nu}\,q_\nu\,F_2(q^2)
	\right]u(p^\prime)\,A_\mu^{\,e}(\vec{q})
	\,,
\label{eq:def:F1F2}
\end{equation}
where $p^\prime=p-q/2$, $p^{\prime\prime}=p+q/2$, 
$q=p^{\prime\prime}-p^\prime$ 
and 
$\sigma^{\mu\nu}=\frac{i}{2}(\gamma^\mu\gamma^\nu-\gamma^\nu\gamma^\mu)$. 
$A_\mu^{\,e}$ is the vector potential of the external static magnetic 
field. 
$F_1$ and $F_2$ are called the charge and magnetic form factors, 
respectively, and 
the charge form factor is normalized so that $F_1(0)=1$. 

The anomalous magnetic moment $a_e$ is the static limit of the 
magnetic form factor $F_2(q^2)$, and it is expressed as
\begin{equation}
	a_e = F_2(0) = Z_2 M
\label{eq:def:ae}
\end{equation}
with
\begin{equation}
	M = \lim_{q^2\to 0} \Tr\bigl(P_\nu(p,q)\,\Gamma^\nu\bigr)\,,
\label{eq:def:M}
\end{equation}
where $Z_2$ is the wave function renormalization constant, 
$\Gamma^\nu$ is the proper vertex part, and 
$P_\nu(p,q)$ is the magnetic projection operator, 
\begin{equation}
	P^\nu(p,q) = \frac{1}{4p^4q^2} 
	\left( \sla{p} - \frac{1}{2}\sla{q} + m \right) 
	\left[ m \gamma^\nu p^2 - \left(m^2+\frac{1}{2}q^2\right)p^\nu \right] 
	\left( \sla{p} + \frac{1}{2}\sla{q} + m \right) 
	\,.
\label{eq:magproj}
\end{equation}
Here, the momentum of incoming lepton $p-\frac{1}{2}q$ 
and that of outgoing lepton $p+\frac{1}{2}q$ 
are on the mass shell so that $p$ and $q$ satisfy 
$p^2 = m^2 - \frac{1}{4}q^2$ and $p\cdot q = 0$. 

We evaluate the anomalous magnetic moment $a_e$ 
in the framework of perturbation theory. 
Because of renormalizability of QED it can be written 
as a power series in $\frac{\alpha}{\pi}$ 
whose coefficients are finite and calculable quantities.

%----------------------------------------------------------------
\subsection{Construction of Feynman parametric integral}
\label{sec:general:integrand}

In perturbative analysis of QED the amplitude is usually 
expressed as an integral of loop momenta flowing through 
the Feynman diagram. 
In this paper we convert it into an integral of Feynman 
parameters $z_i$ assigned to internal lines 
\cite{aldins,Cvitanovic:1974uf}. 

We consider a $2n$th-order lepton vertex diagram $\cal{G}$ 
which describes the scattering of an incoming lepton with 
momentum $p-q/2$ into an outgoing lepton with momentum $p+q/2$ 
by an external magnetic field. 
$\cal{G}$ consists of $2n\!+\!1$ interaction vertices connected by 
$2n$ lepton propagators and $n$ photon propagators, which 
are given in the form (in Feynman gauge):
\begin{equation}
	i \frac{\sla{p}_i+m_i}{p_i^{\ 2}-m_i^{\ 2}},
	\qquad
	\frac{-i g^{\mu\nu}}{p_i^{\ 2} - m_i^{\ 2}} \,,
\label{eq:propagator}
\end{equation}
respectively. 
The momentum $p_i$ may be decomposed as $p_i = k_i + q_i$, 
in which $k_i$ is a linear combination of loop momenta, while 
$q_i$ is a linear combination of external momenta. 
$m_i$ is the mass associated with the line $i$, which is 
temporarily distinguished from each other. 

We introduce an operator $D_i^{\ \mu}$ by 
\begin{equation}
	D_i^{\ \mu} \equiv \frac{1}{2}\int_{m_i^{\ 2}}^{\infty} 
	dm_i^{\ 2}\ \frac{\partial}{\partial q_{i\mu}}
\label{eq:def:operatorD}
\end{equation}
and replace each numerator $\sla{p}_i = \sla{k}_i + \sla{q}_i$ 
of lepton propagators (\ref{eq:propagator}) by $\Sla{D}_i$. 
Since $D_i^{\ \mu}$ does not depend on $k_i$ explicitly, the 
numerators can be pulled out in front of the momentum 
integration as far as the integrand is adequately regularized. 

The product of denominators are combined into one using 
the formula, 
\begin{equation}
	\prod_{i=1}^{N} \frac{1}{\chi_i} 
	= (N-1)! \left[ \prod_{i=1}^{N} \int_{0}^{1}\!dz_i \right]  
	\delta\left(1-\sum_{i=1}^{N}z_i\right)\,
	\frac{1}{\displaystyle\left( \sum_{i=1}^{N} z_i \chi_i \right)^N}
	\,. 
\label{eq:feynmanformula}
\end{equation}
The sum $\sum_i z_i \chi_i$ is a quadratic form of loop momenta 
so that it can be integrated analytically. 
As a consequence the amplitude is converted into an integral 
over Feynman parameters $z_i$ which is expressed 
in a concise form as 
\begin{equation}
	\Gamma_{\cal{G}}^{\ \nu} = 
	\biggl(\frac{1}{4}\biggr)^n (n-1)!\ \mathbb{F}^\nu\!
	\int (dz)_{\cal{G}}\,\frac{1}{U^2 V^n} \,,
\label{eq:def:vertexpart}
\end{equation}
where $N = 3n$ and 
\begin{gather}
	(dz)_{\cal G} = \prod_{i=1}^{N} dz_i\ 
		\delta\left(1-\sum_{i=1}^{N} z_i\right) \,, 
\label{eq:def:dzG} \\
	V = \sum_{i=1}^{N} z_i (m_i^{\ 2} - q_i\cdot Q^\prime_i) \,, 
\label{eq:def:V} \\
	Q^{\prime\ \mu}_i = - \frac{1}{U}\sum_{j=1}^{N}\ 
		q_j^{\ \mu} z_j B^\prime_{ij} \,, 
\label{eq:def:Q} \\
	B^\prime_{ij} = B_{ij} - \delta_{ij}\frac{U}{z_j} \,.
\label{eq:DefBprime}
\end{gather}
In Eq.~(\ref{eq:def:vertexpart}) we have omitted the factor 
$(\alpha/\pi)^n$ for simplicity.
$U$ and $B_{ij}$ are homogeneous polynomials of degree $n$ and $n-1$ in 
Feynman parameters $\{z_i\}$, respectively. 
Their precise definitions are given in later sections. 
The operator $\mathbb{F}^\nu$ is of the form 
\begin{equation}
	\mathbb{F}^\nu = 
	\gamma^{\alpha_1} (\Sla{D}_1 + m_1) \gamma^{\alpha_2} 
	\dots \gamma^\nu \dots 
	\gamma^{\alpha_{2n-1}} (\Sla{D}_{2n} + m_{2n}) \gamma^{\alpha_{2n}} 
	\prod_{k=1}^{n} g_{\alpha_{i_k} \alpha_{j_k}} \,, 
\label{eq:def:operFmu}
\end{equation}
where $\prod_{k} g_{\alpha_{i_k} \alpha_{j_k}}$ is a diagram-specific 
product. 
If $\cal G$ has closed lepton loops $\mathbb{F}^\nu$ also contains 
appropriate trace operations. 

Note that $\mathbb{F}^\nu$ can now be brought into the $z$-integral. 
The operator $D_i^{\ \mu}$ in $\mathbb{F}^\nu$ acts on $1/V^n$ as 
\begin{align}
&	D_i^{\ \mu} \frac{1}{V^n} = \frac{Q^{\prime\,\mu}_i}{V^n} \,, 
\label{eq:actionD} \\
&	D_i^{\ \mu} D_j^{\ \nu} \frac{1}{V^n} 
	= \frac{Q^{\prime\,\mu}_i Q^{\prime\,\nu}_j}{V^n} 
	- \frac{1}{2(n-1)} \frac{g^{\mu\nu} B^{\prime}_{ij}}{UV^{n-1}} \,, 
\label{eq:actionDD} \\
&	D_i^{\ \mu} D_j^{\ \nu} D_k^{\ \rho} \frac{1}{V^n} 
	= \frac{Q^{\prime\,\mu}_i Q^{\prime\,\nu}_j Q^{\prime\,\rho}_k}{V^n} 
\nonumber \\ & \qquad
	- \frac{1}{2(n-1)}
	(g^{\mu\nu} B^{\prime}_{ij} Q^{\prime\,\rho}_{k} 
	+ g^{\nu\rho} B^{\prime}_{jk} Q^{\prime\,\mu}_{i} 
	+ g^{\rho\mu} B^{\prime}_{ki} Q^{\prime\,\nu}_{j}) 
	\frac{1}{UV^{n-1}} \,, 
\label{eq:actionDDD} \\
& \dots \,. \nonumber
\end{align}
\label{eq:contractD}
The result of this operation may be summarized as a set of rules 
for a string of operators 
$D_i^{\ \mu}$:
\begin{itemize}
\item[a)] when $\Sla{D}_i$ and $\Sla{D}_j$ are ``contracted'', 
they are turned into a pair of $\gamma^\mu$ and $\gamma_\mu$ 
times a factor $(-\frac{1}{2}B^\prime_{ij})$.
\item[b)] uncontracted $D_i$ is replaced by $Q^\prime_i$.
\end{itemize}
As a consequence the action of $\mathbb{F}^\nu$ produces a series 
of terms of the form
\begin{equation}
	\mathbb{F}^\nu \frac{1}{U^2 V^n} = 
	\frac{F_0^\nu}{U^2 V^n} + \frac{F_1^\nu}{U^3 V^{n-1}} + \cdots \,,
\label{eq:seriesF}
\end{equation}
where $F_k^\nu$ are polynomials of $B^\prime_{ij}$ and $Q^\prime_i$. 
The subscript $k$ denotes the number of contractions. 
$F_k^\nu$ also includes an overall factor 
$\dfrac{1}{(n-1)(n-2)\cdots(n-k)}$.

\begin{figure}
\includegraphics[scale=1.0]{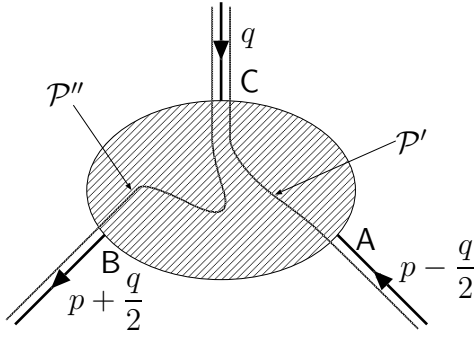}
\caption{A vertex diagram with given external momenta. A choice of 
paths ${\cal P}^\prime$ and ${\cal P}^{\prime\prime}$ are shown.
\label{fig:qpath}}
\end{figure}
It is convenient to replace vectors $Q^{\prime\,\mu}_i$ by scalar 
functions. 
Suppose the momentum 
$p^\mu-\dfrac{q^\mu}{2}$ enters 
the graph $\cal G$ at the point $A$, 
follows the path 
${\cal P}^\prime = {\cal P}(AC)$, and leaves at $C$, and 
$p^\mu+\dfrac{q^\mu}{2}$ enters at $C$, 
follows the path
${\cal P}^{\prime\prime} = {\cal P}(CB)$, and leaves at $A$. 
(Fig.~(\ref{fig:qpath}).)
This can be expressed concisely by 
\begin{equation}
	q_j^{\ \mu} = 
	\eta_{j{\cal P}^\prime} \left(p^\mu-\frac{q^\mu}{2}\right)
	+
	\eta_{j{\cal P}^{\prime\prime}} \left(p^\mu+\frac{q^\mu}{2}\right)\,,
\label{eq:qalongpaths}
\end{equation}
where $\eta_{j{\cal P}^\prime} = (1, -1, 0)$ according to whether 
the line $j$ lies (along, against, outside of) the path ${\cal P}^\prime$. 
Similarly for $\eta_{j{\cal P}^{\prime\prime}}$. 
Substituting Eq.~(\ref{eq:qalongpaths}) in Eq.~(\ref{eq:def:Q}) 
we obtain 
\begin{equation}
	Q^{\prime \mu}_i
	=
	A_i^{\,{\cal P}^\prime} \left(p^\mu-\frac{q^\mu}{2}\right)
	+
	A_i^{\,{\cal P}^{\prime\prime}} \left(p^\mu+\frac{q^\mu}{2}\right) \,,
\label{eq:QinA}
\end{equation}
where
\begin{equation}
	A_i^{\,{\cal P}^{\prime}} = - \frac{1}{U} \sum_{j=1}^{N} 
	\eta_{j{\cal P}^\prime} z_j B^{\prime}_{ji} \,.
\label{eq:AbyBprime}
\end{equation}
Similarly for $A_i^{\,{\cal P}^{\prime\prime}}$. 
$A_i^{\,{\cal P}^{\prime}}$, $A_i^{\,{\cal P}^{\prime\prime}}$ 
will be called scalar currents associated with 
$p^\mu-\dfrac{q^\mu}{2}$, 
$p^\mu+\dfrac{q^\mu}{2}$, respectively. 

If we choose a path ${\cal P} = {\cal P}(AB)$ for $p^\mu$, the 
corresponding scalar current becomes 
$A_i^{\,{\cal P}} = A_i^{\,{\cal P}^\prime} 
+ A_i^{\,{\cal P}^{\prime\prime}}$. 
Note that the choice of ${\cal P}(AB)$ is flexible as far as the 
end points $A$, $B$ are fixed. 
Note also that ${\cal P}(AB)$ no longer depends on $C$.

%----------------------------------------------------------------
\subsection{Building blocks, $B_{ij}$ and $U$}
\label{sec:general:buildingblocks}

In our formalism, the parametric functions $B_{\alpha\beta}$ and $U$ 
provide the basic building blocks which are defined on the chain 
diagram corresponding to the diagram $\cal{G}$. 
Here $\alpha, \beta$ refer to the chains; a chain is a set of 
internal lines that carry the same loop momentum. 
The chain diagram is derived from $\cal{G}$ by amputating all 
the external lines and disregarding the distinction between 
the types of lines. 
Every chain is assumed to be properly directed. 
$B_{\alpha\beta}$ and $U$ are homogeneous polynomials of degree $n-1$ and 
$n$, respectively. They are the quantities that reflect the 
topological structure underlying the diagram $\cal{G}$. 
\begin{figure}
\includegraphics[scale=1.2]{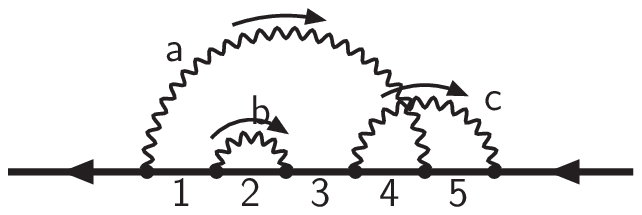}
\includegraphics[scale=0.9]{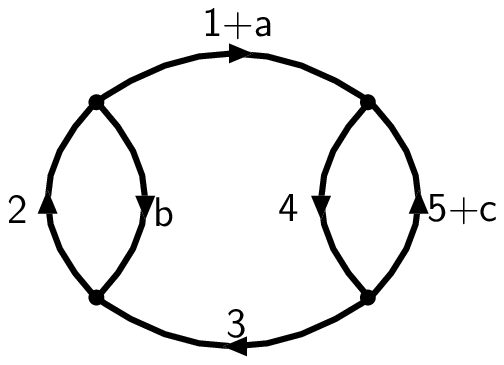}
\caption{A diagram (left) and the chain diagram derived from it (right).}
\end{figure}

$B_{\alpha\beta}$ and $U$ can be obtained recursively by the 
following relations, 
\begin{gather}
	B_{\alpha\beta} 
	= \sum_c \xi_{\alpha,c}\,\xi_{\beta,c}\,U_{{\cal G}/c} \,, 
\label{eq:nakanishiformula} \\
	\xi_{\lambda,s} U 
	= \sum_{\alpha} \xi_{\lambda,s} z_\alpha B_{\lambda\alpha}, 
	\qquad\text{for any $\lambda\in s$} \,, 
\label{eq:ublooplaw}
\end{gather}
starting from $U=\alpha$ for a single loop. 
Here the summation over $c$ runs over all self-nonintersecting 
closed loops on $\cal{G}$. 
The loop matrix $\xi_{\alpha,c}$ is a projector of chain $\alpha$ 
to the loop $c$, which takes $(1, -1, 0)$ according to whether 
$\alpha$ is (along, against, outside of) $c$. 
$U_{{\cal G}/c}$ is the $U$ function for the reduced diagram ${\cal G}/c$ 
that is obtained from $\cal G$ by shrinking the loop $c$ to a point. 
The loop $s$ in Eq.~(\ref{eq:ublooplaw}) is an arbitrary closed loop. 

Alternate and equivalent formulae for $B_{\alpha\beta}$ and $U$ are 
obtained in the following manner. 
Suppose a set of independent self-nonintersecting loops (called 
a fundamental set of circuits) is given and define $U_{st}$ 
by the summation over all chains by 
\begin{equation}
	U_{st} = \sum_{\alpha} z_\alpha\,\xi_{\alpha,s}\,\xi_{\alpha,t} \,,
\label{eq:DefOfUst}
\end{equation}
where $s,t$ are labels of circuits in the set. 
Then, $U$ and $B_{\alpha\beta}$ are given by 
\begin{gather}
	U = \det_{st} U_{st} \,, 
\label{eq:DefOfU} \\
	B_{\alpha\beta} 
	= U \sum_{st} \xi_{\alpha,s}\,\xi_{\beta,t}\,(U^{-1})_{st} \,. 
\label{eq:DefOfB}
\end{gather}
For a given diagram $\cal G$, first we have to identify the 
fundamental set of circuits, and construct the loop matrix 
$\xi_{\alpha,s}$. Then we can obtain $U$ and $B_{\alpha\beta}$ 
according to the formulae above. 

$B_{ij}$ of the lines $i,j$ is identical with $B_{\alpha\beta}$ 
whose indices are such that $i \in \alpha$ and $j \in \beta$. 
$B_{ij}$ satisfies a so-called junction law on each vertex 
if the diagram ${\cal G}$ were regarded as an electric circuit 
in which the Feynman parameter $z_i$ corresponds to the resistance 
of the line $i$: 
\begin{equation}
	\sum_{i} \epsilon_{vi} B_{ij} = 0
\label{eq:junctionlawBij}
\end{equation}
for any vertex $v$ and any internal line $j$,  
where $\epsilon_{vi}$ is called incident matrix defined by
\begin{equation}
	\epsilon_{vi} = 
	\begin{cases}
	1  & \text{if the line $i$ enters the vertex $v$,} \\
	-1 & \text{if the line $i$ leaves the vertex $v$,} \\
	0  & \text{otherwise.}
	\end{cases}
\label{eq:def:incidentmatrix}
\end{equation}
$B_{ij}$ also satisfies a loop-law given by the following relation 
for arbitrary closed loop $s$ and arbitrary line $j$: 
\begin{equation}
	\sum_{i} \xi_{i,s}\,z_i\,B^{\prime}_{ji} = 0 \,.
\label{eq:looplawBij}
\end{equation}
These relations reduce the number of independent elements 
among $B_{ij}$. 
It also provides consistency checks which are useful in the actual 
calculations.

%----------------------------------------------------------------
\subsection{$a_e$ from a set of vertex diagrams summed by 
Ward-Takahashi identity}
\label{sec:general:wt}

A set of vertex diagrams which are derived from a self-energy diagram 
by inserting an external vertex in every lepton propagators 
share many properties. 
Actually we can even go further to relate those integrals to 
a single integral of the self-energy-like diagram through 
the Ward-Takahashi identity.
This relation is useful when we consider higher order calculations 
because it reduces the number of independent integrals substantially. 

It is well known that the proper vertex 
$\Gamma^\mu = \gamma^\mu + \Lambda^\mu$ 
and self-energy part $\Sigma$ are related by the Ward-Takahashi identity
\begin{equation}
	q_\mu \Lambda^\mu 
	= -\Sigma\left(p+\frac{1}{2}q\right) 
	+ \Sigma\left(p-\frac{1}{2}q\right) \,.
\label{eq:wt-exact}
\end{equation}
This relation holds perturbatively as well for $\Sigma_{\cal{G}}$ 
representing the lepton self-energy diagram $\cal{G}$ and the 
sum of vertex diagrams $\Lambda_{\cal{G}}$ that are obtained 
by inserting an external vertex into $\cal{G}$ in every possible way.
Differentiating both sides of Eq.~(\ref{eq:wt-exact}) with respect to 
$q^\mu$ and taking the static limit $q\to 0$ of the external magnetic 
field, we have 
\begin{equation}
	\Lambda^\nu(p,q) 
	\simeq -q^\mu \biggl[
		\frac{\partial\Lambda_\mu(p,q)}{\partial q_\nu}
	\biggr]_{q=0}
	-\frac{\partial\,\Sigma(p)}{\partial p_\nu} \,.
\label{eq:WI}
\end{equation}
We may evaluate $a_e$ starting from either side of 
this expression; a straightforward way is to calculate each 
vertex diagram individually and to gather them up according to 
the left-hand side (\textit{Version B} approach in 
Ref.~\cite{kn1}), 
or else we can combine the set of vertices into one according to 
the right-hand side (\textit{Version A} approach). 
We adopt Version A in the present study. 

In the Feynman parametric form, the $2n$th-order magnetic moment 
associated with a self-energy-like diagram $\cal{G}$ can be 
written as \cite{Cvitanovic:1974um}
\begin{equation}
	M^{(2n)} = \left(\frac{-1}{4}\right)^n (n-1)! 
	\int (dz)_{\cal{G}}\,
	\left[
	\frac{\mathbb{E}+\mathbb{C}}{n-1} \frac{1}{U^2 V^{n-1}}
	+ (\mathbb{N}+\mathbb{Z}) \frac{1}{U^2 V^n}
	\right] \,,
\label{eq:g-2FromSelfEnergy}
\end{equation}
where $\mathbb{E}$, $\mathbb{C}$, $\mathbb{N}$, and $\mathbb{Z}$ 
are a set of operators defined as 
\begin{gather}
	\mathbb{N} = 
	\frac{1}{4}\Tr\bigl[ P_1^\nu p_\nu (2G\mathbb{F}) \bigr] \,, \\
	\mathbb{E} = 
	\frac{1}{4}\Tr\bigl[ P_1^\nu \mathbb{E}_\nu \bigr] \,, \\
	\mathbb{C} = 
	\frac{1}{4}\Tr\bigl[ P_2^{\mu\nu} \mathbb{C}_{\mu\nu} \bigr] \,, \\
	\mathbb{Z} = 
	\frac{1}{4}\Tr\bigl[ P_2^{\mu\nu} \mathbb{Z}_{\mu\nu} \bigr] \,.
\end{gather}
The magnetic projectors $P_1^\nu$ and $P_2^{\mu\nu}$ are derived 
from Eq.~(\ref{eq:magproj}) by averaging over the direction of $q_\mu$, 
and take the following forms:
\begin{gather}
	P_1^{\nu} = \frac{1}{3}\gamma^\nu 
	- \left( 1 + \frac{4}{3}\frac{\sla{p}}{m} \right) 
	\frac{p^\nu}{m} \,, \\
	P_2^{\mu\nu} = \frac{1}{3}\left( 1 + \frac{\sla{p}}{m} \right) 
	\left( g^{\mu\nu} - \gamma^\mu\gamma^\nu 
	+ \frac{p^\mu}{m}\gamma^\nu - \frac{p^\nu}{m}\gamma^\mu \right)\,.
\end{gather}

The operator $\mathbb{F}$ is the numerator part of the self-energy-like 
diagram $\cal{G}$ constructed in the similar form as 
Eq.~(\ref{eq:def:operFmu}): 
\begin{equation}
	\mathbb{F} = 
	\gamma^{\alpha_1} (\Sla{D}_1 + m_1) \gamma^{\alpha_2} \dots 
	\gamma^{\alpha_{2n-1}} (\Sla{D}_{2n-1} + m_{2n-1}) 
	\gamma^{\alpha_{2n}} 
	\prod_{k=1}^{n} g_{\alpha_{i_k}\alpha_{j_k}} \,, 
\label{eq:def:operF}
\end{equation}
which may contain appropriate trace operations if $\cal G$ has 
closed lepton loops. 
The operator $\mathbb{E}^\nu$ is defined by 
\begin{equation}
	\mathbb{E}^\nu = 
	\frac{\partial\,\mathbb{F}}{\partial p_\nu} = 
	\sum_{\rm all\ leptons} A_i \mathbb{F}_i^{\ \nu} \,,
\end{equation}
in which $\mathbb{F}_i^{\ \nu}$ is obtained from $\mathbb{F}$ by 
substituting in the $i$th line:
\begin{equation}
	(\Sla{D}_i + m_i) \to \gamma^\nu \,.
\end{equation}
The operator $\mathbb{Z}^{\mu\nu}$ is defined by 
\begin{equation}
	\mathbb{Z}^{\mu\nu} = 
	\sum_{j} \mathbb{Z}_j^{\ \mu\nu} \,.
\end{equation}
The sum runs only over the lepton lines into which the external 
photon line can be inserted. 
$\mathbb{Z}_j^{\ \mu\nu}$ is obtained from $\mathbb{F}$ by 
substituting in the $j$th line:
\begin{equation}
	(\Sla{D}_j + m_j) \to 
	\frac{1}{2} \bigl[
	\gamma^\mu \gamma^\nu (\Sla{D}_j + m_j) 
	- (\Sla{D}_j + m_j) \gamma^\nu \gamma^\mu \bigr] \,.
\end{equation}
The operator $\mathbb{C}^{\mu\nu}$ is defined by 
\begin{equation}
	\mathbb{C}^{\mu\nu} = 
	\sum_{i<j} C_{ij} \mathbb{F}_{ij}^{\ \mu\nu} \,,
\end{equation}
where $i$ and $j$ refer to all lepton lines. 
$C_{ij}$ is given by 
\begin{equation}
	C_{ij} = \frac{1}{U^2} \sum_{k<l} z_k z_l 
	( B^\prime_{ik} B^\prime_{jl}
	- B^\prime_{il} B^\prime_{jk} ) \,,
\label{eq:defCij}
\end{equation}
where $k$, $l$ are taken from the lepton lines that belong 
to the path on which the momentum $q^\nu$ of the external magnetic 
field flows. 
$\mathbb{F}_{ij}^{\ \mu\nu}$ is obtained from $\mathbb{F}$ by 
substituting in the $i$th and $j$th lepton lines:
\begin{equation}
	(\Sla{D}_i + m_i),\ (\Sla{D}_j + m_j) \to 
	\gamma^\mu,\ \gamma^\nu \,.
\label{eq:replace:DiDj}
\end{equation}
$G$ is given by 
\begin{equation}
	G = \sum_{i} z_i\,A_i \,, 
\label{eq:def:G}
\end{equation}
where the summation runs over the lepton lines on which the 
external momentum $p^\mu$ flows (depending on the choice of 
path ${\cal P}(AB)$ for the scalar currents).

We can now construct the integrand in the following two steps. 
\begin{enumerate}[(I)]
\item\label{item:steps:1} 
Express the integrand as a function of symbols $B_{ij}$, $A_i$, 
$U$, $V$, and $C_{ij}$. 
\item\label{item:steps:2} 
Express those building blocks explicitly in terms of the Feynman 
parameters $z_i$. 
\end{enumerate}

Step~(\ref{item:steps:1}) can be achieved analytically by 
algebraic manipulation programs such as FORM \cite{vermaseren}. 
All the integrals are generated from a small number of 
templates with the permutation of indices according to 
the specific structure of each diagram. 
Step~(\ref{item:steps:2}) is performed along the prescriptions 
outlined above, once $B_{ij}$ and $U$ are 
obtained by the formulae in Section~\ref{sec:general:buildingblocks}.
The magnetic moment contribution (\ref{eq:g-2FromSelfEnergy}) now can be
expressed as a parametric integral: 
\begin{equation}
\begin{aligned}
	M^{(2n)}
	=
	\left(-\frac{1}{4}\right)^{n} (n-1)!\,
	\int (dz)_{\cal G}\,
	&
	\left[
	\frac{1}{n-1} \left(
	\frac{E_0 + C_0}{U^2 V^{n-1}}
	+\frac{E_1 + C_1}{U^3 V^{n-2}}
	+\cdots
	\right)\right. \\
	& \quad
	+	
	\left.\left(
	\frac{N_0 + Z_0}{U^2 V^{n}}
	+\frac{N_1 + Z_1}{U^3 V^{n-1}}
	+\cdots
	\right)\right] \,.
\end{aligned}
\label{eq:g-2FromSelfEnergyExpanded}
\end{equation}

%================================================================

%================================================================
\section{Subtractive UV renormalization procedure}
%================================================================
\label{sec:UVsubtraction}

The amplitude thus far constructed in the previous 
section is divergent in general, 
and the divergences must be removed before carrying out 
the integration numerically. 
The UV divergence arises when one or more loop momenta go 
to infinity. 
This is seen in Feynman parameter space as the parameters $z_i$ 
that belong to loops in a subdiagram go to zero simultaneously. 
It allows power counting rules for identifying the emergence of 
singularities in a similar manner to the ordinary momentum integration. 

We adopt here the subtractive on-shell renormalization. 
In this scheme the renormalization term involving 
an $m$th-order vertex renormalization constant $L_m$ is 
given of the form $-L_m M_{n\!-\!m}$, 
where $M_{n\!-\!m}$ is a $g\!-\!2$ term of order $n\!-\!m$. 
The renormalization constants that appear in QED are 
the mass renormalization constant $\delta m$, 
the wave-function renormalization constant $B$, 
and the vertex renormalization constant $L$. 
They are determined on the mass shell, 
and thus the coupling constant $e$ and lepton's mass $m$ 
are guaranteed to be physical ones. 

To perform renormalization numerically 
our strategy is to prepare the subtraction term 
as an integral over the same domain of integration 
as the original unrenormalized amplitude, and 
to perform point-wise subtraction in which 
singularities of the original integrand are canceled 
point-by-point on the parameter space before the integration. 
To achieve this the renormalization constant $L_m$ and 
the lower-order $g\!-\!2$ term $M_{n\!-\!m}$ are both 
expressed in the parametric integral and combined by the 
Feynman integral formula. 
It is found, however, that the integral is intractable 
if $L_m$ is treated as a whole. 
Instead, we adopt the following two-step intermediate 
renormalization, 
in which $L_m$ is split by 
\begin{equation}
	L_m = L_m^{\ {\rm UV}} + \widetilde{L_m}
	\,,
\label{eq:uv:splitAm}
\end{equation}
and only the UV-divergent part $L_m^{\ {\rm UV}}$ is 
subtracted. 

The subtraction term $L_m^{\ {\rm UV}}\,M_{n\!-\!m}$ is 
found to have a term-by-term correspondence with the 
UV-divergent term of the original integral $M_n$, 
and thus cancels the UV singularities. 
It is identified from the original integrand by simple 
power counting rules. 
This procedure is formulated as \textit{K}-operation. 
The treatment of the UV divergence of self-energy subdiagram 
is slightly more complicated. 
See 
Ref.~\cite{Cvitanovic:1974um,Kinoshita_book} 
and 
Eq.~(\ref{eq:kop:SE}) for details. 

The UV-finite part of the renormalization constant 
is treated separately together with those from other diagrams%
\footnote{
It may contain IR divergences in general and they are 
also subtracted in a similar manner as UV divergences, 
though this subject is not covered here. 
In this article we instead introduce cut-off to treat IR problems. 
}.
This step is called the residual renormalization. 

In this section we shall describe how to construct 
the intermediate renormalization term via \textit{K}-operation. 
It is shown that the subtraction term factorizes exactly into 
the UV-divergent part of the $m$th-order renormalization constant 
and $M_{n\!-\!m}$ by construction. 
This feature is crucial for the subsequent operation 
when the UV divergence arises from more than one divergent 
subdiagrams. 
Such cases are treated more thoroughly in the next section 
in relation to the \textit{forest} structures. 
The factorization property is also significant for the 
residual renormalization step in the sense that the highest 
order of the residual part decreases by two, \textit{e.g.}, 
for the tenth-order diagrams it is sufficient at most 
with the eighth-order terms. 
Therefore the evaluation of the residual part reduces to 
lower-order integrals.

%----------------------------------------------------------------
\subsection{UV divergent subdiagram}
\label{sec:UVsubtraction:subdiagram}

The UV divergence associated with the subdiagram ${\cal S}$ is 
caused by the simultaneous limits $k_i\to\infty$ of all 
loop momenta $k_i$, $i\in{\cal S}$. 
In the parametric representation (\ref{eq:g-2FromSelfEnergyExpanded}) 
this is translated into the 
vanishing of the denominator $U$ at a boundary of Feynman 
parameter space where%
\footnote{
The overall divergence of a self-energy-like diagram drops 
automatically after projecting out the magnetic moment contribution.
}
\begin{equation}
	z_i = 
	\begin{cases}
	{\cal O}(\epsilon) & \qquad i\in{\cal S} \,, \\
	{\cal O}(1)        & \qquad \text{otherwise} \,, 
	\end{cases}
\label{eq:uv:uvlimit}
\end{equation}
with $\epsilon \to 0$. 

To find how a UV divergence arises from a subdiagram ${\cal S}$ 
consisting of $N_{\cal S}$ internal lines and $n_{\cal S}$ loops, 
consider the integration domain (\ref{eq:uv:uvlimit}). 
In the limit $\epsilon \to 0$, the homogeneous polynomials in the 
integrand behave as follows. 
(See Section~\ref{sec:UVsubtraction:buildingblocks} for proofs.)
\begin{equation}
	U = {\cal O}(\epsilon^{n_{\cal S}}),
	\qquad
	V = {\cal O}(1),
\label{eq:uv:scaleUV}
\end{equation}
and 
\begin{equation}
	B_{ij} = 
	\begin{cases}
	{\cal O}(\epsilon^{n_{\cal S}-1}) 
	& \quad \text{if $i,j \in {\cal S}$,} \\
	{\cal O}(\epsilon^{n_{\cal S}}) 
	& \quad \text{otherwise} \quad .
	\end{cases}
\label{eq:uv:scaleBij}
\end{equation}
Let $m_{\cal S}$ be the maximum number of contractions of 
operator $D_i$ within ${\cal S}$. 
Simple power counting shows that the $m$-contracted term of 
$M^{(2n)}$ in Eq.~(\ref{eq:g-2FromSelfEnergyExpanded}) 
is divergent if and only if 
\begin{equation}
	N_{\cal S} - 2 n_{\cal S} \leq \min(m,m_{\cal S}) \,,
\label{eq:uv:conddivergence}
\end{equation}
where $\min(m,m_{\cal S})$ means the lesser of $m$ and $m_{\cal S}$. 
If ${\cal S}$ is a vertex part, 
we have $N_{\cal S} = 3 n_{\cal S}$ and $m_{\cal S} = n_{\cal S}$. 
If ${\cal S}$ is a self-energy part, 
we have $N_{\cal S} = 3 n_{\cal S}-1$ and $m_{\cal S} = n_{\cal S}-1$. 
In both cases Eq.~(\ref{eq:uv:conddivergence}) is satisfied only for 
$m \ge m_{\cal S}$. 
Let us denote the UV limit (\ref{eq:uv:uvlimit}) of $U$ and $B_{ij}$ as
$[U]^S_{UV}$ and $[B_{ij}]_{UV}^S$.

%----------------------------------------------------------------
\subsection{\textit{K}-operation}
\label{sec:UVsubtraction:K-operation}

We are now ready to set up the rules of \textit{K}-operation for 
constructing the intermediate renormalization term. 
Firstly, we summarize our notation. 
${\cal G}/{\cal S}$ denotes a residual diagram which is obtained 
from ${\cal G}$ by shrinking a subdiagram ${\cal S}$ to a point. 
${\cal G}-{\cal S}$ denotes a diagram obtained from ${\cal G}$ 
by eliminating all lines that belong to ${\cal S}$. 

The \textit{K}-operation ${\sf K}_S$ is defined as follows. 
\begin{enumerate}[(1)]
\item \label{item:kstep:maxcontractedterms}
In Eq.~(\ref{eq:g-2FromSelfEnergyExpanded}), 
collect all terms which are maximally contracted within 
the subdiagram $S$.
\item \label{item:kstep:uvlimit}
Replace $U$, $B_{ij}$, $C_{ij}$, and $A_i$ appearing 
in the integrand with their UV-limits, 
$[U]_{\rm UV}^{\cal S}$, 
$[B_{ij}]_{\rm UV}^{\cal S}$, 
$[C_{ij}]_{\rm UV}^{\cal S}$, and 
$[A_i]_{\rm UV}^{\cal S}$, respectively. 
\item \label{item:kstep:replaceV}
Replace $V$ with $V_{\cal S} + V_{{\cal G}/{\cal S}}$, 
where $V_{\cal S}$ and $V_{{\cal G}/{\cal S}}$ are $V$ functions 
of ${\cal S}$ and ${\cal G}/{\cal S}$, respectively. 
\item \label{item:kstep:minussign}
Attach an overall minus sign. 
\end{enumerate}
A na{\"\i}ve UV-limit gives $V \to V_{{\cal G}/{\cal S}}$ 
instead of step (\ref{item:kstep:replaceV}). 
Since $V_{\cal S}$ is a higher order term in $\epsilon$, 
its addition in step (\ref{item:kstep:replaceV}) does not affect the 
UV-limit. 
But it is crucial because it enables us 
to satisfy the exact factorization of the renormalization constant 
and the rest of the amplitude required by the standard renormalization 
\cite{Cvitanovic:1974sv}. 
Furthermore, it enables us to avoid the spurious IR divergence 
which $V_{{\cal G}/{\cal S}}$ alone might develop in other parts 
of the integration domain.

%----------------------------------------------------------------
\subsection{UV-limit of building blocks $U$, $B_{ij}$ and $C_{ij}$}
\label{sec:UVsubtraction:buildingblocks}

Let us now describe step by step how the building blocks of 
the integrand behave in the UV-limit (\ref{eq:uv:uvlimit}). 
It is found that each of them factorizes into two parts, 
one of which depends solely on the subdiagram ${\cal S}$, and 
the other on the residual diagram ${\cal G}/{\cal S}$ alone. 
Since the description given in the literature is somewhat sketchy, 
we shall fill in the gaps here in preparation for automation of 
the procedure. 

The $U$ function is a homogeneous polynomial of Feynman parameters of 
degree $n$ defined by Eq.~(\ref{eq:DefOfU}), which has a simple behavior 
in the limit (\ref{eq:uv:uvlimit}) \cite{Nakanishi:1971} 
\begin{equation}
	[U]_{\rm UV}^{\cal S} = U_{\cal S}\,U_{{\cal G}/{\cal S}}
	\quad 
	\bigl(={\cal O}(\epsilon^{n_{\cal S}}) \bigr).
\label{eq:uvlimit:U}
\end{equation}

In order to obtain the $UV$ limit of $B_{ij}$, 
let us note that for $i\in\alpha$, $j\in\beta$, $B_{\alpha\beta}$ 
of Eq.~(\ref{eq:nakanishiformula}) can be written as 
\begin{equation}
	B_{ij} = \sum_{c}\ \xi_{i,c}\,\xi_{j,c}\,U_{{\cal G}/c} \,.
\label{eq:defBijbyloop}
\end{equation}
Since 
$({\cal G}/c) \cap {\cal S} = {\cal S}/(c \cap {\cal S})$ and 
$({\cal G}/c)/{\cal S} = {\cal G}/(c \cup {\cal S})$, 
the UV-limit of $U_{{\cal G}/c}$ becomes 
\begin{equation}
	[U_{{\cal G}/c}]_{\rm UV}^{\cal S} 
	= 
	U_{{\cal S}/(c \cap {\cal S})} \ 
	U_{{\cal G}/(c \cup {\cal S})} \,.
\label{eq:uvlimit:UoverGc}
\end{equation}
%
% - - - - - - - - - - - - - - - - - - - - - - - - - - - - - - - -
\begin{figure}
\caption{A closed loop $c$ running in ${\cal G}$.
\label{fig:cinG}}
\vskip 2ex
\begin{tabular}{ccc}
\includegraphics[scale=1.0]{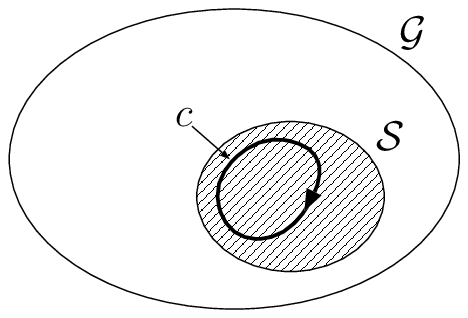} &
\includegraphics[scale=1.0]{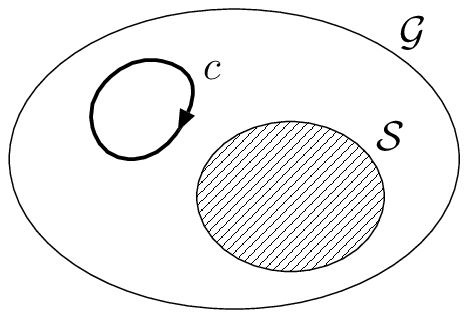} &
\includegraphics[scale=1.0]{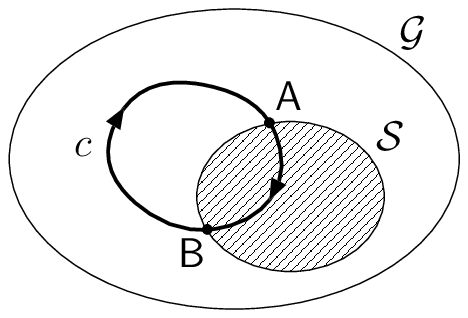} \\
(\ref{case:cands:cinS}) & 
(\ref{case:cands:cinR}) & 
(\ref{case:cands:cinboth}) \\
\end{tabular}
\end{figure}
% - - - - - - - - - - - - - - - - - - - - - - - - - - - - - - - -
%
The explicit form depends on how the loop $c$ runs in ${\cal G}$: 

\begin{enumerate}[{Case }(a){ }]
\item \label{case:cands:cinS}
$c$ is contained in ${\cal S}$. 
(\textit{i.e.}\ $c \subseteq {\cal S}$. ) 
See Fig.~\ref{fig:cinG}~(\ref{case:cands:cinS}).
\\
In this case ${\cal S}/(c \cap {\cal S}) = {\cal S}/c$\ and 
${\cal G}/(c \cup {\cal S}) = {\cal G}/{\cal S}$. 
Therefore 
\begin{equation}
	[U_{{\cal G}/c}]_{\rm UV}^{\cal S} 
	= 
	U_{{\cal S}/c}\,U_{{\cal G}/{\cal S}} 
	\quad 
	\bigl(={\cal O}(\epsilon^{n_{\cal S}-1})\bigr) \,. 
\label{eq:uvlimit:UofcinS}
\end{equation}
The power of $\epsilon$ decreases by 1 since ${\cal S}/c$ has 
one less loops than ${\cal S}$. 
\item \label{case:cands:cinR}
$c$ runs outside of ${\cal S}$. 
(\textit{i.e.}\  $c \subseteq ({\cal G}-{\cal S})$. ) 
See Fig.~\ref{fig:cinG}~(\ref{case:cands:cinR}).
\\
In this case ${\cal S}/(c \cap {\cal S}) = {\cal S}$\ and 
${\cal G}/(c \cup {\cal S}) = ({\cal G}/{\cal S})/c$. 
Therefore
\begin{equation}
	[U_{{\cal G}/c}]_{\rm UV}^{\cal S} 
	= 
	U_{\cal S}\,U_{({\cal G}/{\cal S})/c} 
	\quad 
	\bigl(={\cal O}(\epsilon^{n_{\cal S}})\bigr) \,. 
\label{eq:uvlimit:UofcinR}
\end{equation}
\item \label{case:cands:cinboth}
$c$ is contained in both ${\cal S}$ and ${\cal G}-{\cal S}$. 
(\textit{i.e.}\  $c \cap {\cal S} \neq \emptyset$\ and 
$c \cap ({\cal G}-{\cal S}) \neq \emptyset$. ) 
See Fig.~\ref{fig:cinG}~(\ref{case:cands:cinboth}).
\\
In this case $c \cap {\cal S}$ is an open self-nonintersecting 
path within ${\cal S}$. 
It does not change the number of loops in ${\cal S}$ when the 
path is shrunken to a point. 
Therefore the scaling behavior is 
\begin{equation}
	[U_{{\cal G}/c}]_{\rm UV}^{\cal S} 
	= 
	{\cal O}(\epsilon^{n_{\cal S}}) \,, 
\label{eq:uvlimit:UofcinSandR}
\end{equation}
though the exact factorization does not occur. 
\end{enumerate}

From these observations and Eq.~(\ref{eq:defBijbyloop}) 
we find the following behavior of $B_{ij}$ in the UV limit. 
\begin{enumerate}[I)]
\item \label{case:bijUV:ijinS} 
$B_{ij}$ for $i,j \in {\cal S}$.
\\
The closed loops appearing in the sum in Eq.~(\ref{eq:defBijbyloop}) 
fall into either of the cases 
(\ref{case:cands:cinS}) or (\ref{case:cands:cinboth}), 
the former gives the leading contribution 
whereas the latter does not in the limit 
(\ref{eq:uv:uvlimit}). 
Thus we have 
\begin{equation}
\begin{aligned}
	\left[B_{ij}\right]_{\rm UV}^{\cal S} 
	&= 
	\sum_{c^\prime \subseteq {\cal S}}\ 
	\xi_{i,c^\prime}\,\xi_{j,c^\prime}\,
	U_{{\cal S}/{c^\prime}}\,U_{{\cal G}/{\cal S}} \\
	&=
	B_{ij}^{\cal S}\,U_{{\cal G}/{\cal S}} \,,
\end{aligned}
\label{eq:bijUV:ijinS}
\end{equation}
where the superscript ${\cal S}$ denotes that $B_{ij}^{\cal S}$ 
is the $B$-function defined on the subdiagram ${\cal S}$. 
\item \label{case:bijUV:ijinR} 
$B_{ij}$ for $i,j \in {\cal G}/{\cal S}$.
\\
The closed loops appearing in the sum in (\ref{eq:defBijbyloop}) 
fall into either of the cases 
(\ref{case:cands:cinR}) or (\ref{case:cands:cinboth}), 
both of which give the same order of contributions: 
\begin{equation}
	\left[B_{ij}\right]_{\rm UV}^{\cal S} 
	= 
	\sum_{c^\prime\,\text{in case (\ref{case:cands:cinR})}}\,
	\xi_{i,c^\prime}\,\xi_{j,c^\prime}\,U_{{\cal G}/{c^\prime}}
	+
	\sum_{c^{\prime\prime}\,\text{in case (\ref{case:cands:cinboth})}}\,
	\xi_{i,c^{\prime\prime}}\,\xi_{j,c^{\prime\prime}}\,
	U_{{\cal G}/{c^{\prime\prime}}} \,.
\label{eq:bijUV:ijinR:decomp}
\end{equation}

In the first term on the right-hand side 
the sum over closed loops $c^\prime \subseteq ({\cal G} - {\cal S})$ 
is equivalent to the sum over loops in ${\cal G}/{\cal S} - \{s\}$, 
namely the loops in residual diagram ${\cal G}/{\cal S}$ that 
does not pass through the point $s$, where 
$s$ denotes a point into which the subdiagram ${\cal S}$ has shrunk. 
Therefore, the first term becomes 
\begin{equation}
	U_{\cal S}\,
	\sum_{c^\prime \subseteq ({\cal G}/{\cal S}-\{s\})}\,
	\xi_{i,c^\prime}\,\xi_{j,c^\prime}\,
	U_{({\cal G}/{\cal S})/{c^\prime}} 
	\,.
\label{eq:bijUV:ijinR:1stterm}
\end{equation}

In the second term the closed loop $c^{\prime\prime}$ passing 
through the points $A, B \in {\cal S} \cap ({\cal G}-{\cal S})$ 
is decomposed into two open paths 
${\cal P}(AB) = c^{\prime\prime} \cap {\cal S}$ and 
${\cal P}^\prime (AB) = c^{\prime\prime} \cap ({\cal G} - {\cal S})$. 
The sum over $c^{\prime\prime}$ becomes the sum over a choice 
of points $A, B$ and open paths ${\cal P}(AB)$, ${\cal P}^\prime (AB)$. 
It is shown \cite{Nakanishi:1971} that $U_{{\cal S}/{\cal P}}$ 
satisfies 
\begin{equation}
	U_{\cal S} = \sum_{{\cal P}(AB)}\,U_{{\cal S}/{\cal P}}\,.
\label{eq:nakanishi:UofSoverP}
\end{equation}
On the other hand the path ${\cal P}^\prime (AB)$ becomes a closed 
loop in ${\cal G}/{\cal S}$ that passes through the point $s$ 
to which $S$ has shrunk. 
Thus the second term becomes 
\begin{equation}
	U_{\cal S}\,
	\sum_{c^{\prime\prime} \subseteq {\cal G}/{\cal S},\ c^{\prime\prime}\ni s}\,
	\xi_{i,c^{\prime\prime}}\,\xi_{j,c^{\prime\prime}}\,
	U_{({\cal G}/{\cal S})/{c^{\prime\prime}}} 
	\,.
\label{eq:bijUV:ijinR:2ndterm}
\end{equation}
From Eqs.~(\ref{eq:bijUV:ijinR:1stterm}) and (\ref{eq:bijUV:ijinR:2ndterm}) 
the UV-limit of $B_{ij}$ is 
\begin{equation}
	\left[B_{ij}\right]_{\rm UV}^{\cal S} 
	= 
	B_{ij}^{{\cal G}/{\cal S}}\,U_{\cal S} 
	\qquad
	i,j \in ({\cal G}/{\cal S}).
\label{eq:bijUV:ijinR}
\end{equation}
\item \label{case:bijUV:iinSjinR} 
$B_{mj}$ for $m \in {\cal S}$ and $j \in {\cal G}/{\cal S}$.
\\
This case is relevant only when ${\cal S}$ is a self-energy 
subdiagram, since for the vertex subdiagram case the leading 
contribution comes from the terms in which all lepton lines 
in ${\cal S}$ are contracted with each other. 

We denote the lines which are attached to the subdiagram ${\cal S}$ 
by $i$ and $i^\prime$. (See Fig.~\ref{fig:cinSE}.) 
% - - - - - - - - - - - - - - - - - - - - - - - - - - - - - - - -
\begin{figure}
\caption{A self-energy subdiagram $\cal S$ and a closed loop $c$ 
that passes through $m \in {\cal S}$ and $j \in {\cal G}/{\cal S}$.
\label{fig:cinSE}}
\includegraphics[scale=0.8]{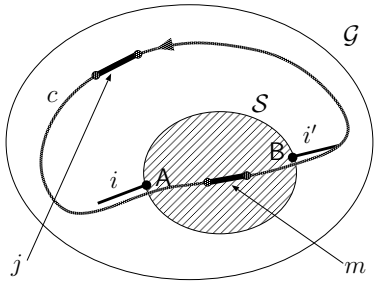}
\end{figure}
% - - - - - - - - - - - - - - - - - - - - - - - - - - - - - - - -
The closed loop $c$ that contains the lines $m\in{\cal S}$ and 
$j\in{\cal G}/{\cal S}$ passes through $i$ and $i^\prime$. 
The sum over loops $c$ is decomposed into the sum over 
${\cal P} = c \cap {\cal G}$ and 
${\cal P}^\prime = c \cap ({\cal G}-{\cal S})$. 
It is shown \cite{Nakanishi:1971} that 
\begin{equation}
	\sum_{{\cal P}}\,\xi_{m,{\cal P}}\,U_{{\cal S}/{\cal P}}
	=
	U_{\cal S}\,A_m^{\cal S} \,,
\label{eq:nakanishi:UofSoverPtoA}
\end{equation}
where $A_m^{\cal S}$ is a scalar current on the line $m$ of the 
diagram ${\cal S}$.
The path ${\cal P}^\prime$ turns into a closed path $c^\prime$ after 
shrinking ${\cal S}$ to a point which passes through the line 
$i \in {\cal G}/{\cal S}$. 
Therefore $B_{mj}$ becomes 
\begin{equation}
\begin{aligned}
	\left[B_{mj}\right]_{\rm UV}^{\cal S}
	&=
	\left(
	\sum_{c^\prime}\,
	\xi_{j,c^\prime},\xi_{i,c^\prime}\,
	U_{({\cal G}/{\cal S})/{c^\prime}}
	\right)\,
	U_{\cal S}\,A_m^{\,\cal S} \\
	&=
	B_{ij}^{{\cal G}/{\cal S}}\,A_m^{\,\cal S}\,U_{\cal S} \,.
\end{aligned}
\label{eq:bijUV:iinSjinR}
\end{equation}
\end{enumerate}

The UV limit of the scalar current $A_j$ follows from 
Eq.~(\ref{eq:AbyBprime}) 
where the path ${\cal P}$ (which replaces $P^\prime$) 
is taken arbitrarily between 
two points attached to external lines. 
We can always choose the path to avoid the line $j$ so that 
$B^\prime_{ij}$ in Eq.~(\ref{eq:AbyBprime}) becomes $B_{ij}$. 

When $\cal S$ is a vertex subdiagram, it is sufficient to 
consider only $A_j$ with $j\in{\cal G}/{\cal S}$, since 
in the leading contributions of the integrand 
all the lepton lines in $\cal S$ are contracted and there is 
no operator $D_i$ left to be turned into scalar current. 
The sum in Eq.~(\ref{eq:AbyBprime}) consists of two parts, 
one from ${\cal P}^\prime = {\cal P}\cap{\cal S}$ and the other 
from ${\cal P}^{\prime\prime} = {\cal P}\cap({\cal G}-{\cal S})$. 
In the limit (\ref{eq:uv:uvlimit}) the scaling behavior 
(\ref{eq:uv:scaleBij}) shows that the former part gives 
sub-leading contribution. 
Therefore, using Eq.~(\ref{eq:bijUV:ijinR}), we obtain 
\begin{equation}
\begin{aligned}
	\left[A_j\right]_{\rm UV}^{\cal S}
	&=
	-\frac{1}{U_{{\cal G}/{\cal S}}}\sum_{i\in{{\cal G}/{\cal S}}}
	\eta_{i{\cal P}^{\prime\prime}}\,z_i\,B_{ij}^{\,{\cal G}/{\cal S}} \\
	&=
	A_j^{\,{\cal G}/{\cal S}} \,.
\end{aligned}
\label{eq:AjUV:vertex:jinR}
\end{equation}

When ${\cal S}$ is a self-energy subdiagram, the scalar currents of 
both $j\in{\cal G}/{\cal S}$ and $j\in{\cal S}$ are relevant. 
\begin{enumerate}[{Case }(a){ }]
\item \label{case:AjUV:SE:jinR} $A_j$ for $j\in{\cal G}/{\cal S}$.
\\
The same argument of the UV limit as in the vertex subdiagram 
applies to this case, which leads to 
\begin{equation}
	\left[A_j\right]_{\rm UV}^{\cal S}
	=
	A_j^{\,{\cal G}/{\cal S}} \,.
\label{eq:AjUV:SE:jinR}
\end{equation}
\item \label{case:AjUV:SE:jinS} $A_m$ for $m\in{\cal S}$.
\\
We choose the path ${\cal P}$ so that it avoids ${\cal S}$. 
Then all $B_{im}$ in (\ref{eq:AbyBprime}) fall into the type 
\ref{case:bijUV:iinSjinR}, whose UV limits are given by 
Eq.~(\ref{eq:bijUV:iinSjinR}). Therefore, 
\begin{equation}
\begin{aligned}
	\left[A_m\right]_{\rm UV}^{\cal S} 
	&=
	-\frac{1}{U_{{\cal G}/{\cal S}}}
	\sum_{k\in{\cal P}} z_k\,B_{ik}^{\,{\cal G}/{\cal S}}
	A_m^{\,{\cal S}} \\
	&=
	A_i^{\,{\cal G}/{\cal S}}\,A_m^{\,{\cal S}} \,,
\end{aligned}
\label{eq:AjUV:SE:jinS}
\end{equation}
where $i$ is the line adjacent to $\cal S$. 
\end{enumerate}

We recall that $C_{ij}$ is derived from the part 
\begin{equation}
	- q^\mu \left[
	\frac{\partial\Lambda_\mu}{\partial q_\nu}
	\right]_{q=0}
\label{eq:WI:vertexpart}
\end{equation}
of Eq.~(\ref{eq:WI}) with the external vertex inserted into 
the line $j$ and differentiated with respect to the 
external momentum $q_\nu$ flowing through the line $i$. 
When ${\cal S}$ is a vertex subdiagram, 
$C_{ij}$ for $i$ or $j$ in ${\cal S}$ have no overall UV divergence, 
since ${\cal S}$ has, effectively speaking, four legs: 
one photon line attached to the external vertex, the other internal 
photon line that is connected to ${\cal G}/{\cal S}$, and 
two internal lepton lines. 
So it is sufficient to consider the cases $i,j \in {\cal G}/{\cal S}$, 
in which the UV-limit of $C_{ij}$ becomes
\begin{equation}
	\left[C_{ij}\right]_{\rm UV}^{\cal S}
	=
	\frac{1}{U_{{\cal G}/{\cal S}}}\,C_{ij}^{\,{\cal G}/{\cal S}}\,.
\label{eq:CijUV:vertex:ijinR}
\end{equation}

When ${\cal S}$ is a self-energy subdiagram, 
the definition (\ref{eq:defCij}) of $C_{ij}$ 
and the UV-limits of $B_{ij}$ lead to the following forms. 
\begin{gather}
	\left[C_{jk}\right]_{\rm UV}^{\cal S}
	=
	\frac{1}{U_{{\cal G}/{\cal S}}}\,C_{jk}^{\,{\cal G}/{\cal S}}
	\qquad
	j,k \in {\cal G}/{\cal S}, \\
	\left[C_{fg}\right]_{\rm UV}^{\cal S}
	=
	\frac{1}{U_{{\cal S}}}\,C_{fg}^{\,\cal S}
	+ 
	\frac{1}{U_{{\cal S}}}\left(
	A_g^{\,{\cal S}} \sum_{h \in {\cal S}}\,z_h B^{\prime\,{\cal S}}_{fh}
	-
	A_f^{\,{\cal S}} \sum_{h \in {\cal S}}\,z_h B^{\prime\,{\cal S}}_{gh}
	\right)\,
	\frac{1}{U_{{\cal G}/{\cal S}}}
	\sum_{j\in{\cal G}/{\cal S}}\,
	z_j B^{\prime\,{\cal G}/{\cal S}}_{ij},
	\qquad
	f,g\in{\cal S}, \\
	\left[C_{fj}\right]_{\rm UV}^{\cal S}
	=
	\frac{1}{U_{{\cal G}/{\cal S}}}\,
	A_f^{\cal S}\,C_{ij}^{\,{\cal G}/{\cal S}}
	+
	\frac{1}{U_{\cal S}}
	\sum_{g\in{\cal S}}\,z_g B^{\prime\,{\cal S}}_{fg}\,
	\frac{1}{U_{{\cal G}/{\cal S}}}
	\sum_{k\in{{\cal G}/{\cal S}}}\,
	z_k B^{\prime\,{{\cal G}/{\cal S}}}_{jk}
	\qquad
	f\in{\cal S}, j\in{{\cal G}/{\cal S}}, 
\end{gather}
where $i$ is the line adjacent to ${\cal S}$.

%----------------------------------------------------------------
\subsection{Factorization property of UV subtraction term}
\label{sec:UVsubtraction:factorization}

Now we proceed to examine the UV subtraction term along the 
steps of \textit{K}-operation to see that it factorizes 
into two parts. 
For simplicity we consider a vertex part $\Gamma_{\cal G}^{\,\nu}$ 
defined in Eqs.~(\ref{eq:def:vertexpart}) and (\ref{eq:seriesF}), 
though the arguments apply to the general cases. 

Suppose the UV divergent subdiagram $\cal S$ is a vertex 
subdiagram. 
In step (1) of \textit{K}-operation we pick up the terms 
which are maximally contracted within ${\cal S}$. 
Such a term among the terms with $k$ contractions, 
$\dfrac{F_k}{U^{2+k} V^{n-k}}$, have the form:
\begin{equation}
	\frac{1}{U^{2}V^{n-k}}\,
	\underbrace{
	\left\{
	\left(\frac{B_{ij}}{U}\right)\cdots
	\right\}
	}_{i,j \in {\cal S}}\,
	\underbrace{
	\left\{
	\left(\frac{B_{i^{\prime}j^{\prime}}}{U}\right)\cdots A_{l^\prime}\cdots
	\right\}
	}_{i^{\prime},j^{\prime},l^{\prime}\in{\cal G}/{\cal S}}
	\,. 
\label{eq:uv:term}
\end{equation}
The first factor in the braces is a product of $B_{ij}$'s with 
$i,j\in{\cal S}$, 
while the second factor is a product that consists of $(k-n_{\cal S})$ 
$B_{i^\prime j^\prime}$'s 
and several scalar currents whose indices $i^\prime,j^\prime$ 
are in ${\cal G}/{\cal S}$. 

In step (2) we consider the UV limit (\ref{eq:uv:uvlimit}). 
It is achieved by replacing the building blocks 
$U$, $B_{ij}$ and $A_j$ 
by their UV limits, 
$[U]_{\rm UV}^{\cal S}$, 
$[B_{ij}]_{\rm UV}^{\cal S}$, 
and $[A_j]_{\rm UV}^{\cal S}$, respectively. 
Then Eq.~(\ref{eq:uv:term}) turns into 
\begin{equation}
\underbrace{
	\frac{1}{U_{{\cal S}}^{\,2}}
	\left\{
	\left(\frac{B_{ij}^{\,{\cal S}}}{U_{\cal S}}\right)\cdots\,
	\right\}
}_{\equiv g[{\cal S}]}
\underbrace{
	\frac{1}{U_{{\cal G}/{\cal S}}^{\,2}}
	\left\{
	\left(\frac{B_{i^{\prime}j^{\prime}}^{\,{\cal G}/{\cal S}}}{U_{{\cal G}/{\cal S}}}\right)\cdots
	A_{l^\prime}^{\,{\cal G}/{\cal S}}\cdots\,
	\right\}
}_{\equiv g[{\cal G}/{\cal S}]}
	\frac{1}{V_{{\cal G}/{\cal S}}^{\,n-k}}
	\,.
\label{eq:uv:termUV}
\end{equation}
The first part depends only on $z_i$ with $i \in {\cal S}$, which we denote 
by $g[{\cal S}]$. 
The second part depends only on $z_i$ with $i \in {\cal G}/{\cal S}$. 
It is denoted similarly by $g[{\cal G}/{\cal S}]$.
In the na{\"\i}ve UV limit $V$ leads to $V_{{\cal G}/{\cal S}}$. 

In step (3) $V_{{\cal G}/{\cal S}}$ is replaced by 
$V_{\cal S} + V_{{\cal G}/{\cal S}}$. 
The integral now becomes
\begin{equation}
	\int (dz)_{\cal G}\,g[{\cal S}]\,g[{\cal G}/{\cal S}]\,
	\frac{1}{\left(V_{\cal S} + V_{{\cal G}/{\cal S}}\right)^{n-k}}
	\,.
\label{eq:uv:integral}
\end{equation}
We shall see that it factorizes into $\cal S$ and 
${\cal G}/{\cal S}$ parts.
Firstly, the identity
\begin{equation}
	1 = 
	\int_0^{1} \frac{ds}{s}\,\delta\left(1-\frac{z_{\cal S}}{s}\right) \ 
	\int_0^{1} \frac{dt}{t}\,\delta\left(1-\frac{z_{{\cal G}/{\cal S}}}{t}\right) \,,
\label{eq:factor:trivial}
\end{equation}
is inserted into the integral, where 
$z_{\cal S}$ and $z_{{\cal G}/{\cal S}}$ are defined by 
$\displaystyle z_{\cal S} = \sum_{i\in{\cal S}} z_i$ and 
$\displaystyle z_{{\cal G}/{\cal S}} = \sum_{i\in{\cal G}/{\cal S}} z_i$, 
respectively. 
Secondly, we rescale the Feynman parameters as follows:
\begin{equation}
\begin{aligned}
	z_i \to s z_i &\qquad i \in {\cal S} \\
	z_i \to t z_i &\qquad i \in {{\cal G}/{\cal S}}
\end{aligned}
\label{eq:factor:scale}
\end{equation}
Since $V$-functions are homogeneous polynomial of degree 1, 
they scale in such a manner as 
$V_{\cal S} \to s\,V_{\cal S}$ and 
$V_{{\cal G}/{\cal S}} \to t\,V_{{\cal G}/{\cal S}}$.
Other parts of the integrand and the integration measure 
also scale accordingly. 

Then using the Feynman integral formula:
\begin{equation}
	\Gamma(k+l) \int_0^{1} ds\,dt\ \delta(1\!-\!s\!-\!t)\,
	\frac{s^{k-1} t^{l-1}}{(sA+tB)^{k+l}}
	=
	\frac{\Gamma(k)}{A^k}\,\frac{\Gamma(l)}{B^l} \,, 
\label{eq:factor:feynmanformula}
\end{equation}
the integral is shown to be factorized into two parts:
\begin{equation}
\begin{aligned}
	&\int dz_{\cal S}\,\delta(1\!-\!z_{\cal S})\,g[{\cal S}] 
	\int dz_{{\cal G}/{\cal S}}\,\delta(1\!-\!z_{{\cal G}/{\cal S}})\,g[{\cal G}/{\cal S}] 
	\int ds\,dt\,\delta(1\!-\!s\!-\!t)\,\frac{s^{\alpha-1}t^{\beta-1}}{\left(sV_{\cal S} + tV_{{\cal G}/{\cal S}}\right)^{\alpha+\beta}} \\
	&\qquad=
	\int (dz)_{\cal S}\,\frac{g[{\cal S}]}{V_{\cal S}^{\ \alpha}}
	\times
	\int (dz)_{{\cal G}/{\cal S}}\,\frac{g[{\cal G}/{\cal S}]}{V_{{\cal G}/{\cal S}}^{\ \beta}} \,,
\end{aligned}
\end{equation}
where $\alpha$ and $\beta$ are constants determined by 
the rescaling (\ref{eq:factor:scale}). 

Based on those observations the whole integral of the vertex part 
$\Gamma_{\cal G}^{\nu}$ is shown to be factorized in UV limit as 
\begin{equation}
	{\sf K}_{\cal S} \Gamma_{\cal G}^{\ \nu} 
	= 
	L^{\rm UV}_{\cal S} \Gamma_{{\cal G}/{\cal S}}^{\ \nu} \,,
\label{eq:kop:vertex}
\end{equation}
where $L^{\rm UV}_{\cal S}$ is the UV divergent part of the vertex 
renormalization constant $L_{\cal S}$ and 
$\Gamma_{{\cal G}/{\cal S}}^{\ \nu}$ is the vertex part of the 
residual diagram ${{\cal G}/{\cal S}}$.

When $\cal S$ is a self-energy subdiagram, the factorization is 
not apparent because not all $\Sla{D}_m$ with $m\in{\cal S}$ are 
contracted. From Eqs.~(\ref{eq:bijUV:iinSjinR}) and (\ref{eq:AjUV:SE:jinS}) 
we can symbolically write the uncontracted $\Sla{D}_m$ as 
\begin{equation}
	\left[\Sla{D}_m\right]_{\rm UV}^{\cal S} 
	= 
	A_m^{\,{\cal S}}\,\Sla{D}_{i^{\prime\prime}}^{\,{{\cal G}/{\cal S}}}
	\,,
\label{eq:opDinSE}
\end{equation}
where $i^{\prime\prime}$ is a fictitious line related to 
$i$ and $i^\prime$.
After a little algebra one finds \cite{Cvitanovic:1974um,Kinoshita_book} 
\begin{equation}
	{\sf K}_{\cal S} \Gamma_{\cal G}^{\ \nu} 
	= 
	{\delta m}^{\rm UV}_{\cal S} \Gamma_{{\cal G}/{\cal S}(i^{*})}^{\ \nu}
	+ 
	B^{\rm UV}_{\cal S} \Gamma_{{\cal G}/{\cal S},i^\prime}^{\ \nu} \,,
\label{eq:kop:SE}
\end{equation}
where ${\delta m}^{\rm UV}_{\cal S}$ is the UV divergent part of the 
mass renormalization constant $\delta m_{\cal S}$ and 
$B^{\rm UV}_{\cal S}$ 
is that of the wave function renormalization constant $B_{\cal S}$. 
${\cal G}/{\cal S}(i^{*})$ denotes the diagram obtained 
by shrinking ${\cal S}$ to a point, where $i^{*}$ indicates 
two-point vertex between lines $i$ and $i^\prime$. 
${\cal G}/{\cal S},i^\prime$ denotes the diagram derived from 
${\cal G}$ by shrinking ${\cal S}$ to a point and eliminating 
the line $i^\prime$.
It can be reduced to the form $\Gamma_{{\cal G}/{\cal S}}^{\ \nu}$ 
after integration by part with respect to $z_{i}$. 

The factorization of \textit{K}-operation is crucial 
when there are more than one subdiagrams that cause UV divergences, 
since this property guarantees that the successive operation of 
${\sf K}_{{\cal S}_k}$ is consistent with the forest structure. 

%----------------------------------------------------------------

%================================================================

%================================================================
\section{Forest formula}
%================================================================
\label{sec:forest_formula}

A Feynman diagram that appears at higher-order terms of 
perturbation theory may have complicated UV-divergence structures. 
In many textbooks they are treated in a recursive formulation 
so that the inner subdivergences of a renormalization part 
should be subtracted prior to the subtraction of its own overall 
divergence. 
It is natural in the framework of renormalization theory, 
for it is derived from the requirement 
that the divergences should be resolved by local counterterms. 
It is also tractable in general for hand manipulations since 
the subtractions are performed step by step from lower order parts 
and the number of steps are, as it turns out, relatively small. 
It is, however, not so convenient in our numerical approach 
in which the singularities due to the UV divergences 
are canceled point-by-point in the Feynman parameter space. 
To achieve this we have to prepare the subtraction terms as 
integrals defined in the same parameter space as that of 
the original unrenormalized amplitude. 

An explicit solution of the recursive formulation is given 
by Zimmermann's forest formula \cite{Zimmermann:1969jj}. 
Each source of the UV-divergence is related to a \textit{forest}, 
a set of UV-divergent subdiagrams, and the subtraction term 
associated with the forest is constructed by the subtraction 
operations for the subdiagrams applied successively 
to the unrenormalized amplitude. 
The whole subtraction terms are generated along the complete 
set of forests. 

In our numerical approach the subtraction operation is 
given by \textit{K}-operation for a single subdivergence. 
As seen in the previous section it retains the factorization 
property, which guarantees the successive \textit{K}-operations. 
Therefore, once a UV-divergent structure is known in the form 
of a forest, we can obtain the integrand of the subtraction term 
\cite{Cvitanovic:1974um}. 
Although the subtraction scheme presented here is identical with that 
developed in Ref.~\cite{Cvitanovic:1974um}, it is more readily adaptable 
for code generation. 

The forest formula is much more useful for the automated scheme. 
The forests are given by the combinations of non-overlapping 
subdiagrams. 
So the complete identification of UV-divergent structures is 
obtained by purely combinatorial procedure from the set of 
all UV-divergent subdiagrams. 
Thus it is readily implemented in terms of forests, 
which enables us to obtain fully UV-renormalized amplitude of 
a diagram $\cal G$.

%----------------------------------------------------------------
\subsection{Definition of forests}

We begin with the inclusion relations between subdiagrams. 
When two subdiagrams ${\cal S}_i$ and ${\cal S}_j$ share 
no vertices nor lines they are called \textit{disjoint}. 
When all lines of ${\cal S}_i$ belong to ${\cal S}_j$, 
the subdiagram ${\cal S}_i$ is \textit{included} in 
the subdiagram ${\cal S}_j$. 
In this case ${\cal S}_i$ and ${\cal S}_j$ are called 
\textit{nested}. 
Otherwise they are called \textit{overlapping}, in which 
${\cal S}_i$ and ${\cal S}_j$ share some vertices and 
lines while one is not included in the other. 

A forest is defined as a set of subdiagrams whose elements satisfy 
the condition that any pairs of them are disjoint or nested. 
(The empty set is also allowed.) 
When a forest contains the diagram $\cal G$ itself, it is called 
\textit{full} forest. 
Otherwise it is \textit{normal} forest. 
For the calculation of $g\!-\!2$ term it is sufficient to consider 
only normal forests. 
We denote the set of all possible normal forests of a diagram 
$\cal G$ by $\mathfrak{F}({\cal G})$. 

%----------------------------------------------------------------
\subsection{Forest formula and \textit{K}-operations}

Assume $\mathfrak{C}_{\cal S}$ is a subtraction operator 
associated with a subdiagram $\cal S$. 
The renormalized amplitude $M^\prime_{\cal G}$ of a diagram 
$\cal G$ is obtained from the unrenormalized amplitude $M_{\cal G}$ 
by \textit{forest formula} \cite{Zimmermann:1969jj}: 
\begin{equation}
	M^\prime_{\cal G} = \sum_{f\in\mathfrak{F}({\cal G})}
	\left[
		\prod_{{\cal S}_i\in f} (-\mathfrak{C}_{{\cal S}_i}) 
	\right]
	M_{\cal G}
	\,.
\label{eq:forest:formula}
\end{equation}
Here the sum is taken over all forests $f$ of the diagram $\cal G$. 
The order of operation in the product is arranged so that 
the inner subdiagrams are applied first. 

In our approach the subtraction operation is provided 
as \textit{K}-operation for performing 
the intermediate renormalization in numerical procedure. 
Recall that the integrand of the subtraction term obtained 
by \textit{K}-operation factorizes exactly into the UV-divergent 
part of renormalization constant and the lower-order $g\!-\!2$ term 
by construction. 
This feature enables us to apply repeatedly \textit{K}-operations 
if there is another UV-divergent subdiagram in the forest.

%----------------------------------------------------------------
\subsection{Procedure}

The subtraction term associated with a forest is obtained 
by the successive \textit{K}-operations. 
The concrete procedure is given as follows. 
Suppose the forest $f$ consists of $m$ subdiagrams, 
$f = \{{\cal S}_1, \dots, {\cal S}_m\}$. 
They are arranged in such an order that the inner subdiagrams 
come ahead. 

The scaling like Eq.~(\ref{eq:uv:uvlimit}) for a forest $f$ 
can be defined similarly by introducing the scaling parameters 
$\left\{\epsilon_k\right\}_{k = 1,\,\cdots,\,m}$ as 
\begin{equation}
	z_i = O(\epsilon_k) \qquad i \in {\cal S}_k,
\label{eq:forest:uvlimit}
\end{equation}
where ${\cal S}_k$ is the inner-most subdiagram of the forest $f$ 
which contains the line $i$. 
We define the UV limit 
$\left[g(\{z_j\})\right]^{f}_{\rm UV}$ 
of a function $g(\{z_j\})$ of Feynman parameters 
as the leading term of $g(\{z_j\})$ 
in the successive limits 
\begin{equation} 
	\left[g(\{z_j\})\right]^{f}_{\rm UV} 
	= 
	\lim_{\epsilon_{m} \rightarrow 0} 
	\cdots 
	\lim_{\epsilon_1 \rightarrow 0} 
	g(\{z_j\}) \, . 
\label{eq:UVLimit_for_a_forest} 
\end{equation} 

Due to the factorization property 
we can construct the subtraction term corresponding to the 
forest $f$ by the repeated applications of \textit{K}-operations. 
The \textit{K}-operation for $k$th subdiagram ${\cal S}_k$ 
is applied to the reduced diagram 
${\cal G}/({\cal S}_{1} \cup \cdots \cup {\cal S}_{k\!-\!1})$ 
which has been obtained by shrinking the subdiagrams up to 
$(k-1)$th subdiagrams to points. 
The integrand of the subtraction term is obtained in the following way. 
\begin{enumerate}[(1)]
\item 
In Eq.~(\ref{eq:g-2FromSelfEnergyExpanded}) 
collect all terms which are maximally contracted within 
the subdiagram ${\cal S}_i$ for $i=1,\dots,m$.
\item 
Replace $U$, $B_{ij}$, $C_{ij}$ and $A_j$ 
appearing in the integrand by their UV limits, 
$\left[U\right]_{\rm UV}^{f}$, 
$\left[B_{ij}\right]_{\rm UV}^{f}$, 
$\left[C_{ij}\right]_{\rm UV}^{f}$, 
and 
$\left[A_j\right]_{\rm UV}^{f}$, 
respectively.
\item 
Arrange $V$ in the limit to take the form 
\begin{equation} 
	V_{\widetilde{{\cal S}_1}} 
	+ \cdots 
	+ V_{\widetilde{{\cal S}_m}} 
	+ V_{{\cal G}/({\cal S}_1 \cup \cdots \cup {\cal S}_m)}
	\,,
\label{eq:forest:replaceV}
\end{equation} 
where $\widetilde{{\cal S}_k}$ is a subdiagram obtained from ${\cal S}_k$ 
by shrinking all inner subdiagrams to points. 
\item 
Attach the overall sign $(-1)^m$. 
\end{enumerate}

The end result of this construction is identical with what 
was obtained in Refs.~\cite{kl1,Cvitanovic:1974um,Kinoshita_book}.
The advantage of the forest approach is that 
it is readily translatable into computer code, 
which of course is crucial for automation of our entire formulation. 

%================================================================

%================================================================
\section{q-Type Feynman diagrams}
%================================================================
\label{sec:q-type}

In this section and the next we focus on the particular type 
of diagrams which have no closed lepton loops. 
We call such a diagram as \textit{q-type}. 
The q-type diagram has a simple structure, which allows simple 
identification of various graph-theoretical notions embedded 
in the diagram. 
The key features relevant for automated scheme of calculations 
may be listed as follows:
\begin{enumerate}[(a)]
\item\label{item:feature:a} 
systematic generation of diagrams are easily done. 
\item\label{item:feature:b} 
a set of independent loops are easily identified. 
\item\label{item:feature:c} 
subdiagrams relevant for the UV divergence are easily 
identified.
\end{enumerate}
They enable us to develop efficient algorithms and implementations. 

In this section the features (\ref{item:feature:a}) 
and (\ref{item:feature:b}) are discussed. 
A set of algorithms to obtain the complete set of topologically 
independent diagrams is presented. 
The feature (\ref{item:feature:b}) is related to 
the construction of the topological forms $B_{ij}$ and $U$, 
which provide the building blocks of the integrand. 
The feature (\ref{item:feature:c}) will be discussed in the 
next section in relation to the subtraction of UV divergences.

%----------------------------------------------------------------
\subsection{Definition and diagram representation}
\label{sec:q-type:defs}

\begin{figure}
\caption{An example of q-type diagram.\label{fig:qtype}}
\includegraphics[scale=1.0]{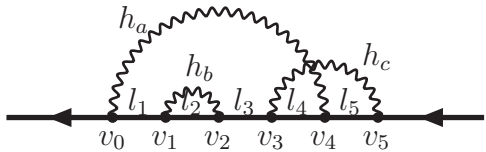}
\end{figure}

A q-type self-energy-like diagram of $2n$th order is given 
by a path ${\cal P}$ consisting of lepton lines emanating from 
the incoming lepton $\bar{\psi}(y)$ and terminating at the 
outgoing lepton $\psi(x)$, with $n$ photon lines attached 
to the path at their both ends; there is no closed lepton loop. 
A typical diagram is shown in Fig.~\ref{fig:qtype}.
In QED there is only one type of interaction, the coupling of 
electromagnetic current $j^\mu = \bar{\psi}\gamma^\mu\psi$ to the 
gauge potential $A_\mu$, which is represented in Feynman rules as 
a trivalent vertex, at which a photon line is attached to the 
lepton line path.
Here we consider only one-particle irreducible (1PI) diagrams.

We denote $2n$ vertices as $v_j$ $( j=0,\dots,(2n-1) )$, and 
adopt the following convention throughout the rest of the paper. 
Every q-type diagram is drawn in such a way that the path 
${\cal P}$ passes from the right to the left. 
The vertices $v_j$ sequentially lie on the path ${\cal P}$ as 
$v_0\,\dots,v_{2n-1}$ from the left to the right. 
A lepton line is denoted as $l_k$ $(k=1,\dots,(2n-1))$ which runs 
from a vertex $v_k$ to another vertex $v_{k-1}$. 
A photon line which connects two vertices $v_{i_k}$ and $v_{j_k}$, 
is denoted as $h_k$ $(k=a,b,\dots)$, where label $k$ is taken to be 
an alphabet. The photon line is also represented by the pair of 
two endpoints $(v_{i_k}, v_{j_k})$. 
For later convenience, a direction of the photon line is chosen 
by $v_{i_k}\rightarrow v_{j_k}$ where the indices are ordered 
as $i_k < j_k$.
We also denote the lines collectively as $\{l\}$. 

A q-type diagram ${\cal G}$ is uniquely specified by the set of 
$n$ photon lines, {\em i.e.}, the set of pairs of vertices as
\begin{equation}
	{\cal G} = \left\{
		(v_{i_1}, v_{j_1}), \dots, (v_{i_{n}}, v_{j_{n}})
		\right\} \,,
\end{equation}
where $i_1,j_1,\dots,i_{n},j_{n}$ take values in 
$\{0,\dots,(2n-1)\}$ exclusively.
To avoid the ambiguity of representation, we further impose the 
following condition: 
\begin{equation}
	\begin{aligned}
	i_k < j_k, 
	&\quad\text{for each pair}\ (v_{i_k},v_{j_k}), \\
	i_k < i_{k^\prime}, 
	&\quad\text{between two pairs}\ k < k^\prime \,.
	\end{aligned}
\label{eq:diag_repr_cond}
\end{equation}

%----------------------------------------------------------------
\subsection{Circuits and loop matrix}
\label{sec:q-type:circuits}

Consider the chain diagram associated with a q-type diagram.
A {\em circuit} of the graph is a self-nonintersecting closed 
path along the graph. 
The maximal independent set of them defines a fundamental set of 
circuits, which provides a complete basis of closed loops of the 
graph.
The fundamental set of circuits of a graph is crucial in the 
calculation of $U$ and $B_{ij}$. 

In the case of q-type diagram, a path ${\cal C}_r$ composed of 
a photon line $h_r = (v_{i_r}, v_{j_r})$ and lepton lines 
$l_{i_r+1},\dots,l_{j_r}$ that lie between two endpoints of 
the photon line form a circuit. 
The direction of the circuit can be taken naturally along those of 
lepton and photon lines. 
The fundamental set is thus chosen by the set of paths 
$\left\{{\cal C}_r\right\}_{r=1,\dots,n}$.

The loop matrix $\xi_{j,{\cal C}_r}$ is given almost trivially. 
Once $\xi_{j,{\cal C}_r}$ is known, $U$ and $B_{ij}$ are calculated 
according to the formula 
(\ref{eq:DefOfUst}), (\ref{eq:DefOfU}), and (\ref{eq:DefOfB})
given in Section~\ref{sec:general}, 
or 
(\ref{eq:new_formula_for_U}), (\ref{eq:def_of_quantity_A}), 
and (\ref{eq:new_formular_for_Bfunction}) 
in Appendix~\ref{sec:detail}.

%----------------------------------------------------------------
\subsection{Time-reversal symmetry}
\label{sec:q-type:time-reversal}

By using the time-reversal symmetry of QED, we can further relate 
distinct diagrams to each other, to reduce independent set of 
diagrams.
Two q-type Feynman diagrams are equivalent in time-reversal and 
give rise to the same contribution to anomalous magnetic moment 
if and only if they are the images of each other under the reversal 
of the directions of all lepton lines in ${\cal P}$.

For a time-reversal equivalent pair of diagrams, it is sufficient 
to evaluate either of the two. 
The asymmetric diagrams in time-reversal dominate at higher orders, 
which implies that the number of distinct diagrams to be evaluated 
is cut down to almost half by considering time-reversal symmetry.

%----------------------------------------------------------------
\subsection{Algorithms}
\label{sec:q-type:algorithms}

Suppose $2n$ vertices are placed on a lepton line path ${\cal P}$. 
Then the complete set of topologically distinct q-type diagrams of 
$2n$th order is obtained as follows:
\begin{quote}
\begin{itemize}
\item[Step 1.] 
Connect a pair of vertices by a photon line in every possible way.
\item[Step 2.] 
Pick only 1PI diagrams and discard others.
\item[Step 3.] 
Drop either of the pair of equivalent diagrams under time-reversal. 
\end{itemize}
\end{quote}

Step 1 is a process to list up all possible ways to make $n$ pairs 
out of the vertices $\{0,\dots,(2n-1)\}$.
A procedure to make $k$ pairs from $2k$ elements is given as 
follows; the elements are assumed to be ordered in a line.
\begin{quote}
Pick an element at the left end of the line, and another one from the 
rest to form a pair. Repeat the process to the remaining $2k-2$ 
elements to make $k-1$ pairs until there is no element left.
\end{quote}
By considering $2k-1$ ways to make a pair, we can generate all 
possible pairings recursively.
The total number of ways is $(2k-1)!!$.

Next we go to step 2. 
The diagram corresponding to a pairing generated above may not 
satisfy the 1PI condition. 
Since a q-type diagram is connected by lepton lines, it is 
sufficient to check whether it stays connected when one of 
the lepton lines is eliminated.
\begin{quote}
A q-type diagram is 1PI if and only if for each lepton line $l_k$, 
there exists at least one photon line 
that {\em steps over} the lepton line, {\em i.e.}, two end points of 
photon line, $(v_i, v_j)$, satisfy $v_i \leq v_{k-1}$ and $v_j \geq v_k$
simultaneously.
\end{quote}
The diagrams that do not match the condition shall be discarded. 

The time-reversal operation in step 3 is done by substituting the 
index $k$ of vertex $v_k$ by $2n-1-k$. 
A q-type diagram ${\cal G}$ is mapped to ${\cal G}^\prime$ by the 
substitution of indices followed by the reshuffling of pairs to 
satisfy the conventions (\ref{eq:diag_repr_cond}).
If ${\cal G}$ is invariant under the time-reversal, it should 
be kept with the symmetry factor one. 
Otherwise, either of ${\cal G}$ or ${\cal G}^\prime$ should be 
kept with the symmetry factor two; we adopt the rule that 
the diagram ${\cal G}$ is chosen when the lexicographical order 
of the patterns of indices representing the diagrams  
${\cal G}$ is ahead of ${\cal G}^\prime$.

%----------------------------------------------------------------
\subsection{Number of diagrams}
\label{sec:q-type:number}

Based on the above consideration, the number of 1PI q-type diagrams 
${\cal N}_{n}$ of $n$ loops is given recursively by the following 
relation (disregarding time-reversal symmetry):
\begin{equation}
	{\cal N}_{n} = (2n-1)!! 
	- \sum_{(k_1,\dots,k_m) \in {\mathfrak P}_{n}}
	\prod_j\,{\cal N}_{k_j} \,,
\end{equation}
where ${\mathfrak P}_{n}$ denotes the set of {\em ordered} partitions 
of $n$ ({\em i.e.} $(1,2)$ and $(2,1)$ should be distinguished). 

Table~\ref{table:num_s-t_EC} shows the number $N_{n}$ of 
independent q-type diagrams for $n \leq 7$ 
as well as that of symmetric ones and 
that of asymmetric ones under time-reversal. 
(${\cal N}_n = N_{n}^{\phantom{n}\text{sym}} + 2 N_{n}^{\phantom{n}\text{asym}}$.)
\begin{table}[tbh]  
\caption{%
The number of independent 1PI diagrams of q-type, $N_{n}$, with $n$ 
loops.}
\begin{center} 
\begin{tabular}{c|r|r|r} 
\hline \hline 
\makebox[3em]{$n$} & 
\makebox[6em]{$N_{n}$} & 
\makebox[6em]{$N_{n}^{\phantom{n}\text{sym}}$}  & 
\makebox[6em]{$N_{n}^{\phantom{n}\text{asym}}$} \\ 
\hline 
 1 & 1 & 1 & 0 \\ 
 2 & 2 & 2 & 0 \\ 
 3 & 8 & 6 & 2 \\ 
 4 & 47 & 20 & 27 \\ 
 5 & 389 & 72 & 317 \\ 
 6 & 4226 & 290 & 3936  \\ 
 7 & 55804 & 1198 & 54606 \\ 
\hline \hline 
\end{tabular} 
\end{center} 
\label{table:num_s-t_EC}
\end{table} 
It also demonstrates that the incorporation of time-reversal 
symmetry efficiently reduces the number of independent 
diagrams to be evaluated 
$\left( N_{n}^{\phantom{n}\text{sym}} + N_{n}^{\phantom{n}\text{asym}} \right)$
at higher orders.

%================================================================
\section{UV divergence structure of q-type diagrams}
%================================================================
\label{sec:q-type_UV}

For a given Feynman diagram, it is required in the renormalization 
process to pick up all the 1PI subdiagrams that have {\em overall} 
ultraviolet (UV) divergence. 
They are referred to as UV divergent 1PI subdiagrams. 
By the power counting of superficial divergence, there are two 
types of UV divergent subdiagrams in q-type diagrams in QED, 
namely, the lepton self-energy-like subdiagram, and the vertex 
subdiagram. 
Every subtraction term in the subtractive renormalization procedure 
corresponds to the Zimmermann's forest, a combination of subdiagrams 
whose loop momenta go to infinity.

For a q-type diagrams, the above prescription is implemented quit 
simply, reflecting the graph theoretical properties of the 
diagram. 
In this section we discuss the UV structure of the q-type diagrams 
and describe an algorithm to compose subdiagrams and forests of 
the diagram.

%----------------------------------------------------------------
\subsection{UV divergent subdiagrams}
\label{sec:q-type_UV:subdiagrams}

A subdiagram relevant for the UV divergence is either 
of the self-energy type or of the vertex type. 
For a q-type diagram in which all vertices are located on 
the lepton line path $\cal P$ 
a subset of vertices are denoted by one or more segments 
of the path. 
Thus a subdiagram of these types of a q-type diagram 
corresponds to a single segment of the path, and 
it is specified by the indices of two end-point vertices. 

Therefore, to obtain all the divergent subdiagrams of a q-type 
diagram we have only to find every possible pair of indices 
$[i,j]$, $0 \le i < j \le (2n-1)$ that satisfies the following 
two conditions: 
\begin{quote}
The subdiagram corresponding to the segment $[i,j]$ is classified into 
the self-energy-like type or the vertex type. 
The number of `floating' photon line (only either one of the 
two endpoints of the photon line lies on the segment $[i,j]$) is 
zero (for self-energy-like subdiagram) or one (for vertex subdiagram).
\end{quote}
And, 
\begin{quote}
It is one-particle irreducible, {\em i.e.}, it stays connected 
when any one of the lepton lines that belong to the subdiagram 
is eliminated. 
\end{quote}
The second condition is satisfied when for every lepton line, 
$l_{i+1},\dots,l_j$, there is at least one photon line that 
belongs to the subdiagram (both endpoints of it lie between 
$v_i$ and $v_j$) which {\em steps over} the lepton line. 
The photon line $h_k = (v_{i_k}, v_{j_k})$ {\em steps over} 
the lepton line $l_s = (v_{s-1}, v_s)$ when 
$i \le i_k \le s-1$ and $s \le j_k \le j$ simultaneously.

It is noted that subdiagrams of q-type diagrams also do not 
contain lepton loops. 
The residual diagram of a q-type diagram which is obtained 
by shrinking the subdiagram to a point is again of q-type. 
Recall that it is related to lower-order $g-2$ term. 
The UV subtraction procedure is closed within the q-type diagrams. 

%----------------------------------------------------------------
\subsection{Forests}
\label{sec:q-type_UV:forests}

To begin with, we define the inclusion relation between two 
subdiagrams, $S_a$ and $S_b$: 
\begin{description}
\item[disjoint] if $S_a$ and $S_b$ do not share any vertices 
nor lines, 
{\em i.e.} $S_a \cap S_b = \emptyset$.
\item[overlapping] if $S_a$ and $S_b$ share some vertices and 
lines though one is not completely included in the other. 
\item[nested] if $S_a$ (or $S_b$) is a subset of the other, 
{\em i.e.} $S_a \subset S_b$ or $S_a \supset S_b$.
\end{description}

For the q-type diagrams, the inclusion relation is mapped to that 
of two segments. 
The relation between two subdiagrams represented by 
$S_a = [i_a, j_a]$  and $S_b = [i_b, j_b]$ 
is one of the following (assuming that $i_a \leq i_b$):

\begin{tabular}{ll}
\makebox[.45\textwidth][l]{\textit{disjoint}\ \ if $j_a < i_b$,} & 
\begin{picture}(100,20)(0,0)
\put(0,0){\thicklines\line(1,0){100}}
\put(10,0){\circle*{2}}
\put(30,0){\circle*{2}}
\put(70,0){\circle*{2}}
\put(90,0){\circle*{2}}
\put(10,0){\line(0,1){10}}
\put(10,10){\line(1,0){20}}
\put(30,10){\line(0,-1){10}}
\put(8,-10){$i_a$}
\put(28,-10){$j_a$}
\put(70,0){\line(0,1){10}}
\put(70,10){\line(1,0){20}}
\put(90,10){\line(0,-1){10}}
\put(68,-10){$i_b$}
\put(88,-10){$j_b$}
\end{picture}
\\
\textit{overlapping}\ \ if $i_a < i_b \leq j_a < j_b$, &
\begin{picture}(100,30)(0,0)
\put(0,0){\thicklines\line(1,0){100}}
\put(10,0){\circle*{2}}
\put(40,0){\circle*{2}}
\put(60,0){\circle*{2}}
\put(90,0){\circle*{2}}
\put(10,0){\line(0,1){10}}
\put(10,10){\line(1,0){50}}
\put(60,10){\line(0,-1){10}}
\put(8,-10){$i_a$}
\put(62,7){$j_a$}
\put(40,0){\line(0,-1){10}}
\put(40,-10){\line(1,0){50}}
\put(90,-10){\line(0,1){10}}
\put(35,-17){$i_b$}
\put(88,-17){$j_b$}
\end{picture}
\\
\textit{nested}\ \ if $i_a \leq i_b$ and $j_b \leq j_a$. &
\begin{picture}(100,40)(0,0)
\put(0,0){\thicklines\line(1,0){100}}
\put(10,0){\circle*{2}}
\put(30,0){\circle*{2}}
\put(70,0){\circle*{2}}
\put(90,0){\circle*{2}}
\put(10,0){\line(0,1){15}}
\put(10,15){\line(1,0){80}}
\put(90,15){\line(0,-1){15}}
\put(8,-10){$i_a$}
\put(88,-10){$j_a$}
\put(30,0){\line(0,1){10}}
\put(30,10){\line(1,0){40}}
\put(70,10){\line(0,-1){10}}
\put(28,-10){$i_b$}
\put(68,-10){$j_b$}
\end{picture}
\\
\end{tabular}
\vskip 4ex
%
% - - - - - - - - - - - - - - - - - - - - - - - - - - - - - - - -
\begin{figure}
\caption{A forest composed of nested subdiagrams 
$\gamma_1,\dots,\gamma_4$ (left), and 
the corresponding cascade structure (right).}
\begin{center}
\begin{minipage}{.45\textwidth}
\includegraphics[scale=.7]{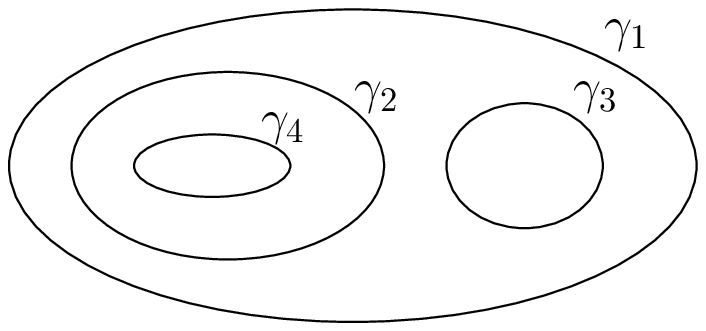}
\end{minipage}
\begin{minipage}{.3\textwidth}
\includegraphics[scale=.7]{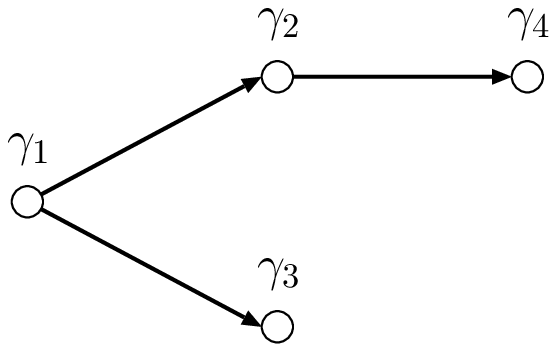}
\end{minipage}
\end{center}
\end{figure}
% - - - - - - - - - - - - - - - - - - - - - - - - - - - - - - - -

A forest is defined as such a set of subdiagrams that 
any two of its elements are not overlapping with each other; 
they are disjoint or nested.
Since we are currently interested in the magnetic form factor, 
it is sufficient to consider only the `normal' forests which 
do not contain the diagram $\cal G$ itself. 
On the other hand, a forest that contains $\cal G$ is called 
`full' forest.

A complete set of forests of a diagram is generated by finding 
all the combinations of the subdiagrams, and discarding those 
which contain the overlapping subdiagrams. (This particular 
procedure is not restricted to the q-type diagrams.)
A cascade structure of subdiagrams of a forest is reproduced 
by referring to the inclusion relation between subdiagrams. 
This information is required during the subtraction process 
which is performed for the divergence corresponding to the 
inner subdiagrams first. 

% - - - - - - - - - - - - - - - - - - - - - - - - - - - - - - - -
%----------------------------------------------------------------

%================================================================
\section{Automated flow of calculation}
%================================================================
\label{sec:flow}

\begin{figure}[t]
\caption{%
Flow of process to generate the numerical integration codes 
from the diagram representation.
\label{fig:flow}}
\begin{center}
\includegraphics[scale=0.7]{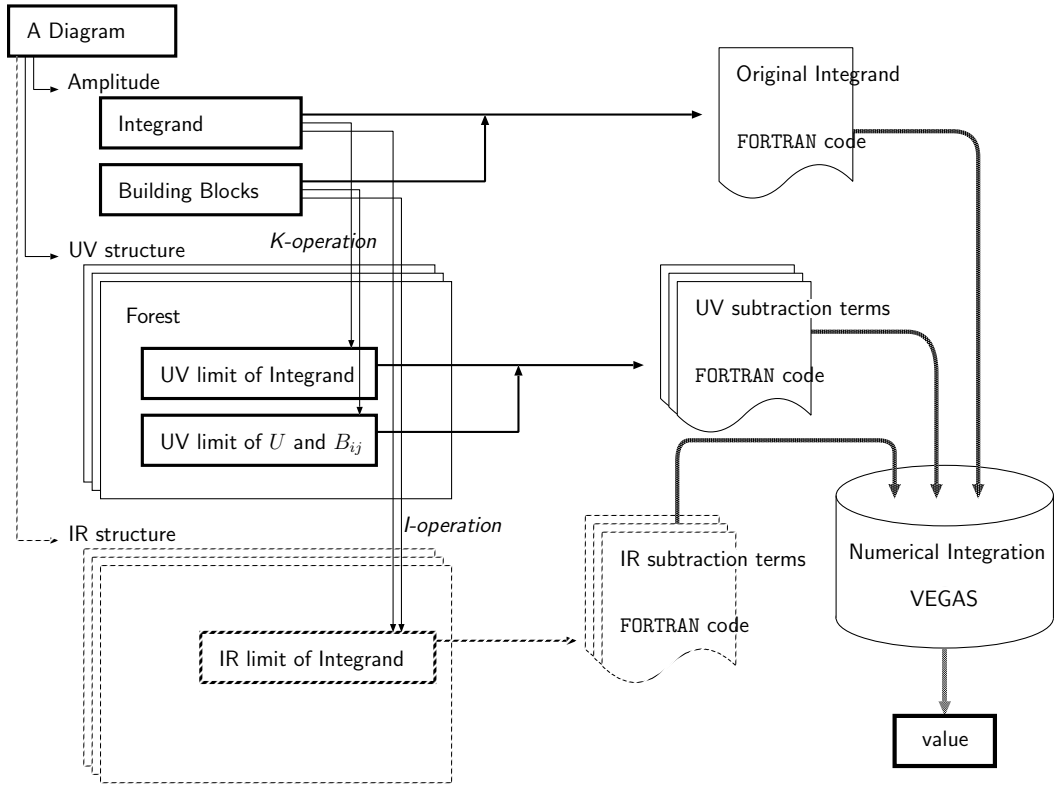}
\end{center}
\end{figure}

In this section we present the flow of the process to generate the 
numerical integration code for evaluating an individual diagram 
from its representation 
indicated by the rectangular box at the upper-left corner of 
Fig.~\ref{fig:flow}. 
The provided information enables us to construct 
the amplitude in the form of Feynman parametric 
integrals in terms of building blocks, $U$, $B_{ij}$, 
scalar currents, and so forth. 
This follows exactly the pattern developed for the sixth- and 
eighth-order cases \cite{Cvitanovic:1974uf,kl1}. 
Next, the ultraviolet divergence is treated via \textit{K}-operation 
which identifies and subtracts the most divergent part of the 
original integral, corresponding to a specific UV limit. 
The treatment of the whole divergence structure is organized by 
the Zimmermann's forest formula. 
The infrared divergence remaining in the individual diagram should also 
be subtracted away, though this article does not cover this subject. 
Finally, the (intermediate-) renormalized amplitude constructed from 
the original amplitude and the set of subtraction terms 
is turned into a FORTRAN code, which is readily processed by the 
numerical integration system such as VEGAS
\cite{lepage}, 
an adaptive Monte-Carlo integration routine.

%----------------------------------------------------------------
\subsection{Diagram generation}
\label{sec:flow:diagram}

We begin by generating a complete set of topologically distinct 
q-type diagrams of a given order according to the algorithm 
in Section~\ref{sec:q-type:algorithms}. 
The implementation of the algorithm is achieved in C++.
Each diagram is expressed by a \textit{single-line} representation 
which describes the pattern of pairings of vertices 
by the photon propagators. 
The diagrams are then named after a certain convention and 
stored in a plain text file. 
All subsequent steps refer to this file for the diagram data. 

Names and forms of all relevant diagrams of 
the sixth- 
and the eighth-orders 
are listed in Refs.~\cite{kn2,Kinoshita_book}. 
We adopt the following convention for the tenth-order diagrams. 
The W-T summed diagrams are classified into two groups, 
one of which is time-reversal symmetric (72 diagrams) 
and the other is asymmetric (317 diagrams). 
They are sorted in a lexicographical order within each group, 
and then given serial numbers with a prefix ``X'' 
which stands for the tenth-order in a Roman numeral, 
first for the symmetric ones (\texttt{X001}, \dots, \texttt{X072}), 
then for the asymmetric ones (\texttt{X073}, \dots, \texttt{X389}).

%----------------------------------------------------------------
\subsection{Subdiagram search and forest construction}
\label{sec:flow:subdiagram}

All UV divergent subdiagrams of a q-type diagram are 
identified according to the algorithms 
in Section~\ref{sec:q-type_UV:subdiagrams}. 
Then all forests are constructed according to the description 
in Section~\ref{sec:q-type_UV:forests} 
by generating every possible combination of 
subdiagrams which does not contain any overlapping pairs. 
The inclusion relation between subdiagrams is examined 
prior to this step. 

The cascade structure among the subdiagrams of a forest is 
also identified and stored in a tree form. 
It is, however, not mandatory 
since the order of successive subtraction operations of a forest 
is automatically respected if we adopt the regulation that 
the diagrams of smaller sizes are processed first. 
It is because the sizes of subdiagrams $S_i$ and $S_j$ satisfy 
$S_i < S_j$ whenever $S_i$ is contained in $S_j$.

The implementation is carried out in both Perl and C++. 
To demonstrate the fast algorithms and implementations 
we generated the diagrams and their forests up to 14th order, 
which took less than 10 minutes on an ordinary PC.

%----------------------------------------------------------------
\subsection{Constructing unrenormalized integrand}
\label{sec:flow:integrand}

A single-line representation of a q-type diagram may be directly 
translated into the form of unrenormalized integrand. 
Recall that the integrand $\mathbb{F}$ of a q-type diagram is 
given by Eq.~(\ref{eq:def:operF}): 
\begin{equation}
	\mathbb{F} = 
	\gamma^{\mu_1}\left(\Sla{D}_1 + m_1\right)\gamma^{\mu_2}
	\cdots
	\gamma^{\mu_{2n\!-\!1}}
	\left(\Sla{D}_{2n\!-\!1} + m_{2n\!-\!1}\right)\gamma^{\mu_{2n}}\,
	\times
	\prod_{k=1}^{n}\,g_{\mu_{i_k}\mu_{j_k}}
	\,,
\label{eq:F:qtype}
\end{equation}
where the diagram-specific product $\prod g_{\mu_{i_k}\mu_{j_k}}$ 
in this case is determined by the pairing pattern 
\begin{equation}
	\left\{
	(v_{i_1}, v_{j_1}),\,(v_{i_2}, v_{j_2}), \dots
	\right\}
	\,.
\label{eq:vertexparing}
\end{equation}
The basic form of integrand is common to all q-type diagrams, 
where we have only to make permutation of indices to construct 
the integrand of a particular diagram. 
Therefore we can make full use of a template to perform this step. 

We implemented this step as a Perl program, which translates 
the single-line representation of the diagram into an explicit 
form of $\mathbb{F}$ and put it into a template of FORM program. 
It is then processed by FORM to perform 
the analytic integration over all loop momenta, 
the trace calculation, 
contractions of $D$ operators, 
and other algebraic manipulations, 
which yields the unrenormalized integrand expressed as 
a polynomial of the building blocks, 
$U$, $B_{ij}$, $C_{ij}$ and $A_j$, 
integrated over Feynman parameters. 
This follows exactly the method developed for the sixth- and 
eighth-order cases \cite{Cvitanovic:1974uf,kl1}. 
The output is in the FORM-readable form for the subsequent steps of 
deriving various UV limits, as well as in the form of FORTRAN code.

%----------------------------------------------------------------
\subsection{Constructing building blocks}
\label{sec:flow:buildingblocks}

The building blocks of integrand, $U$ and $B_{ij}$, 
are determined from the underlying topological structure of 
the diagram called chain diagram 
\cite{Cvitanovic:1974uf,Kinoshita_book}. 
First we identify chains and chain variables 
$z_\alpha = \sum_{i\in\alpha} z_i$. 
The fundamental set of circuits $\{{\cal C}_r\}_{r=1,\dots,n}$ 
of the diagram is identified according to the specification 
in Section~\ref{sec:q-type:circuits}, and 
the loop matrix $\xi_{\alpha,{\cal C}_r}$ is constructed accordingly. 
Once $\xi_{\alpha,{\cal C}_r}$ is known the building blocks 
$U$, $B_{ij}$ are obtained as homogeneous polynomials of $\{z_\alpha\}$ 
by Eqs.~(\ref{eq:DefOfUst}), (\ref{eq:DefOfU}), and (\ref{eq:DefOfB}). 
Other building blocks, 
$\widetilde{C}_{ij}$ $(C_{ij} = \widetilde{C}_{ij}/U)$, $A_j$, 
and $V$ 
are also constructed in terms of $U$, $B_{ij}$ and Feynman 
parameters $\{z_i\}$ 
by Eqs.~(\ref{eq:defCij}), (\ref{eq:AbyBprime}), (\ref{eq:def:V}), 
and (\ref{eq:def:G}): 
\begin{gather}
	A_j = \frac{1}{U}\sum_{i} z_i\,B^{\prime}_{ij} \,,
\label{eq:qtype:Aj} \\
	V = \sum_{i} z_i - G \,,
\qquad	
	G = \sum_{i} z_i\,A_i \,,
\label{eq:qtype:V}
\end{gather}
where the sum runs over all lepton lines. 

The calculation of $U$, $B_{ij}$ and $\widetilde{C}_{ij}$ 
involves algebraic manipulations such as determinants and cofactors 
of matrices whose elements are polynomials of Feynman parameters 
$\{z_\alpha\}$. 
We implemented this step in three ways.
\begin{enumerate}[i)]
\item \label{item:ubij:bymaple} 
The algebraic manipulations are performed by MAPLE. 
We identify the loop matrix from the diagram representation and 
prepare a MAPLE program to calculate $U$, $B_{ij}$, and $C_{ij}$ 
according to 
Eqs.~(\ref{eq:DefOfUst}), (\ref{eq:DefOfU}), (\ref{eq:DefOfB}), 
and (\ref{eq:defCij}). 
The output is in the FORM-readable form for the subsequent operations. 
The FORTRAN code is also generated from it via FORM. 

\item \label{item:ubij:byformulae} 
We developed concise formulae, 
(\ref{eq:new_formula_for_U}) and (\ref{eq:def_of_quantity_A}), 
to provide the coefficients 
of the $U$-function directly from the loop matrix. 
The coefficients $u_{p_1 p_2 \dots p_m}$ are defined by 
\begin{equation}
	U = \sum_{\{p_i\}}\,u_{p_1 p_2 \dots p_m}\,
	z_{\alpha_1}^{\,p_1}\,z_{\alpha_1}^{\,p_1}
	\dots
	z_{\alpha_n}^{\,p_m}\,,
\label{eq:ubij:coeffU}
\end{equation}
for every possible combination of $\{p_i\}$, where 
$p_i$ takes $0$ or $1$ and $\displaystyle\sum_{i=1}^{m} p_i = n$. 
$m$ is the total number of chains. 
Similarly for $B_{ij}$ and $C_{ij}$ by the formulae 
described in Appendix~\ref{sec:detail}. 
They are implemented in C++.

\item \label{item:ubij:bylibrary} 
The algebraic manipulations are also handled in C++ 
by constructing proper data structures (or \textit{classes}). 
We developed a simple polynomial class and implemented the 
calculation of $U$ and $B_{ij}$ according to 
Eqs.~(\ref{eq:DefOfUst}), (\ref{eq:DefOfU}), and (\ref{eq:DefOfB}). 
We also developed another version with the help of 
\texttt{GiNaC} \cite{ginac}, 
an algebraic manipulation library in C++.

\end{enumerate}
The current version of automation system relies on the implementation 
\ref{item:ubij:bymaple}) for no particular reason. 
The scalar currents $A_j$ and the function $V$ are constructed 
from $U$ and $B_{ij}$ according to 
Eqs.~(\ref{eq:qtype:Aj}) and (\ref{eq:qtype:V}).

%----------------------------------------------------------------
\subsection{Constructing UV subtraction terms}
\label{sec:flow:uv}

The UV divergences of a diagram are identified as Zimmermann's 
forests which are the combinations of UV divergent subdiagrams. 
We construct a set of UV subtraction terms, each of which 
corresponds to a particular forest, by the successive applications 
of \textit{K}-operations to the unrenormalized integrand. 

The subtraction term is constructed by the following three steps. 
\begin{quote}
\begin{enumerate}[1.]
\item \label{item:uvstep:1}
Find the UV limit of the building blocks.
\item \label{item:uvstep:2}
Find the UV limit of the integrand.
\item \label{item:uvstep:3}
Modify the UV limit of $V$-function in the denominator 
to satisfy the factorizability requirement of subtraction terms. 
\end{enumerate}
\end{quote}
Those steps are achieved by simple power counting applied 
to the original (unrenormalized) integrand and building blocks, 
without referring to any lower-order constructs. 
The whole implementation is done in Perl with the help of 
MAPLE and FORM for symbolic manipulations. 
This follows exactly the scheme developed for the sixth- 
and eighth-order cases \cite{Cvitanovic:1974uf,kl1}. 

% - - - - - - - - - - - - - - - - - - - - - - - - - - - - - - - -
\subsubsection{UV limit of building blocks}
\label{sec:flow:uv:buildingblocks}

The explicit form of building blocks, $U$, $B_{ij}$, and 
$\widetilde{C}_{ij}$, in the UV limit (\ref{eq:uv:uvlimit}) 
related to a subdiagram ${\cal S}$ is given as the leading term 
in power series expansion by $\epsilon$ under the rescaling 
of the Feynman parameters as 
\begin{equation}
	z_i \to \epsilon\,z_i \,,
	\qquad
	i\in{\cal S} \,.
\label{eq:uvstep:rescale}
\end{equation}
The procedure is implemented as a MAPLE or FORM program, in which 
the scaling rules (\ref{eq:uvstep:rescale}) are generated 
from the information of the subdiagram. 

For a forest consisting of more than one subdiagrams 
the above procedure is successively applied with each 
subdiagram ${\cal S}_k$. 
The order of operations is determined referring to 
the cascade structure of the forest 
so that the inner subdiagrams are applied first. 

The UV limit of scalar current $A_j$ is constructed from the 
UV limits of $U$ and $B_{ij}$.

% - - - - - - - - - - - - - - - - - - - - - - - - - - - - - - - -
\subsubsection{UV limit of the integrand}
\label{sec:flow:uv:integrand}

According to the formulation in 
Section~\ref{sec:UVsubtraction:subdiagram}, 
the UV divergent part of the integrand in the UV limit 
(\ref{eq:uv:uvlimit}) is derived from 
the most contracted terms within the subdiagram $\cal S$. 
This part is simply extracted by counting the number of $B_{ij}$ 
with $i,j \in {\cal S}$ in each term of the unrenormalized 
integrand. 

The procedure is implemented as a FORM program, which reads 
the expression of integrand constructed in 
Section~\ref{sec:flow:integrand} 
and picks up the terms which are the products of 
the specified number of $B_{ij}$ whose indices belong to the 
subdiagram ${\cal S}$. 
For a forest consisting of more than one subdiagrams, 
the above counting are applied successively according 
to the order with the inner subdiagrams first. 
The FORM program is generated referring to the forest data. 

% - - - - - - - - - - - - - - - - - - - - - - - - - - - - - - - -
\subsubsection{UV limit of $V$-function in the denominator}
\label{sec:flow:uv:replaceV}

The UV limit of $V$-function in the denominator 
$V_{{\cal G}/{\cal S}}$ is replaced as follows: 
\begin{equation}
	V_{{\cal G}/{\cal S}} \to V_{\cal S} + V_{{\cal G}/{\cal S}}
\label{eq:uvstep:replaceV}
\end{equation}
in the step (3) of \textit{K}-operation to guarantee the 
factorization property. 
At a glance this operation might require the explicit construction 
of $V$-functions of lower order diagrams, 
$V_{\cal S}$ and $V_{{\cal G}/{\cal S}}$ individually. 
However, it turns out that this can be achieved by adopting the 
following rule: 
in the construction of $V$ by Eq.~(\ref{eq:qtype:V}) the scalar 
currents $A_j$ with $j\in {\cal S}$ should be replaced by 
$\left[A_j\right]_{\cal S}$, where 
$\left[A_j\right]_{\cal S}$ is given by 
\begin{enumerate}[(a)]
\item dropping all terms containing $B_{ij}$ with $i\in{\cal S}$ 
and $j\in{{\cal G}/{\cal S}}$, 
\item replacing other $B_{ij}$ and $U$ by their UV limits.
\end{enumerate}
Thus the replacement of $V$-function is also accomplished 
solely by the limit operations from the original building blocks.

%----------------------------------------------------------------
\subsection{Symbolic expressions of subtraction terms}
\label{sec:flow:symbolic}

The subtraction term has a symbolic expression in terms of 
the product of renormalization constants and lower order 
magnetic moment part, each term of which is related to 
the particular structure of the corresponding forest. 
The identification of the symbolic expression is 
achieved by pattern matching based on the rule set. 

We prepare the rule set for recognizing the particular 
pattern of subdiagrams (after shrinking the inner 
subdiagrams to points) 
and identify the form of the expressions. 
The whole implementation is done in Perl and the rule 
set is also generated automatically from the basic 
representation of the self-energy-like diagrams. 

The symbolic expression is better suited for human recognition. 
It will also be relevant when we perform the residual 
renormalization.

%----------------------------------------------------------------
\subsection{Controlling the whole steps}
\label{sec:flow:control}

Each step of code generation is achieved by separate 
Perl programs with the help of MAPLE and FORM, while the 
flow of the whole process is governed by a shell script. 
It takes the name of the diagram as an input and performs 
the following operations:
\begin{enumerate}[a)]
\item Pick out the corresponding expression of the 
diagram from data file.
\item Construct each component of the integration code.
\item Gather up the FORTRAN codes in the end.
\end{enumerate}
The whole process of code generation for each W-T summed diagram of 
tenth order takes 10--20 minutes on an ordinary PC.

%----------------------------------------------------------------
% - - - - - - - - - - - - - - - - - - - - - - - - - - - - - - - -
%----------------------------------------------------------------

%================================================================
\section{Conclusion and Discussions}
%================================================================
\label{sec:conclusion}

In this paper we presented an automated scheme of code generation 
for evaluating higher order QED corrections of the lepton anomalous 
magnetic moment by means of numerical method. 
We constructed an algorithm and concrete procedure to obtain 
UV-finite amplitudes for a particular set of diagrams without 
lepton loops, which we call q-type diagrams. 
Our current concern is the tenth-order corrections, though, the 
scheme itself is applicable to an arbitrary order. 

We implemented our procedure as a set of Perl programs with the 
help of symbolic manipulation systems, FORM and MAPLE. 
From a single-line representation of a diagram it generates 
numerical integration codes in FORTRAN, which are ready to 
be processed by VEGAS, an adaptive Monte-Carlo integration 
routine. 

The programs have been tested for lower-order diagrams and 
confirmed that they reproduce the codes for the sixth order 
and eighth order diagrams previously constructed. 
They are now being applied to tenth-order diagrams. 
At present, the diagrams which have only vertex renormalization 
were processed and test runs were performed. 
Those diagrams corresponds to 2232 vertex diagrams among 
6354 q-type diagrams of tenth-order.
Crude evaluation showed no sign of divergent behavior, 
which confirms that our scheme is working well. 
They are currently put to production runs. 

The remaining 4122 diagrams have not only UV divergent 
self-energy subdiagrams but also infrared (IR) divergences. 
The simplest way to deal with the IR problem is 
to give a small mass $\lambda$ to photons, which requires 
no further work on the automating code, and is being pursued 
as the first step. 

This scheme has thus far been tested successfully with the 
sixth-order q-type diagrams, and reproduced the analytic 
result after proper treatment of residual renormalization terms. 
For the eighth-order q-type diagrams 
it seems successful so far except for a few diagrams which 
suffer from more severe IR divergences than logarithmic. 
One remedy is to subtract full part of renormalization term 
by taking properly into account the effect of self-mass term 
of the leptons. 
This modification will be implemented within a slight extension 
to the current automation code and is being worked out. 

To obtain a result independent of $\lambda$ it is necessary 
to incorporate IR subtraction terms in a manner similar to 
that of UV counterterms. 
The subsequent step of residual renormalization is 
our next issue.

Our scheme has been elucidated for the q-type diagrams 
in this paper, though the formalism is not restricted to 
that type of diagrams. 
The practical algorithm for the construction of building 
blocks, that are related to the underlying topological structure 
of the diagram, and the identification of UV divergent subdiagrams 
rely on the particular properties of the q-types. 
However, they can be extended to incorporate more general cases. 
We will then have a fully automatic scheme for evaluating QED 
diagrams of lepton $g\!-\!2$.

%================================================================

\begin{acknowledgments}
M.~H.'s work is supported in part by Ministry of Education, 
Science and Culture of Japan, 
Grant-in-Aid for Scientific Research (15740173, 13135223). 

T.~K.'s work is supported by the National Science Foundation 
under Grant No. PHY-0098631. 
T.~K. thanks the Eminent Scientist Invitation Program of RIKEN,
Japan, for the hospitality extended to him where a part
of this work was carried out.
T.~K. is also supported during his stay in Japan 
by Ministry of Education, Science and Culture of Japan, 
Grant-in-Aid for Scientific Research on Priority Areas, 13134101.

M.~N.'s work is partly supported by Ministry of Education, 
Science and Culture of Japan, 
Grant-in-Aid for Scientific Research (C) 15540303, 2003-2005.

The numerical calculation has been performed on the RIKEN 
Super Combined Cluster System (RSCC).
\end{acknowledgments}

%================================================================
\appendix

%================================================================
\section{Classification of Diagrams contributing to $A_1^{(10)}$}
%================================================================
\label{sec:classification} 

There are 12672 vertex-type Feynman diagrams at the tenth order. 
We classify them into six sets according to the type of virtual 
lepton loop(s) and how they appear in a Feynman diagram. 
Every figure in this appendix should be supplied with one external 
vertex in all non-trivial places of the internal lepton lines. 

All diagrams in the sets I, II, III and IV are obtained by inserting 
vacuum-polarization and/or light-by-light-scattering 
subdiagrams of appropriate orders 
into lower-order q-type diagrams. 

\begin{figure}[htb] 
\includegraphics[scale=0.8]{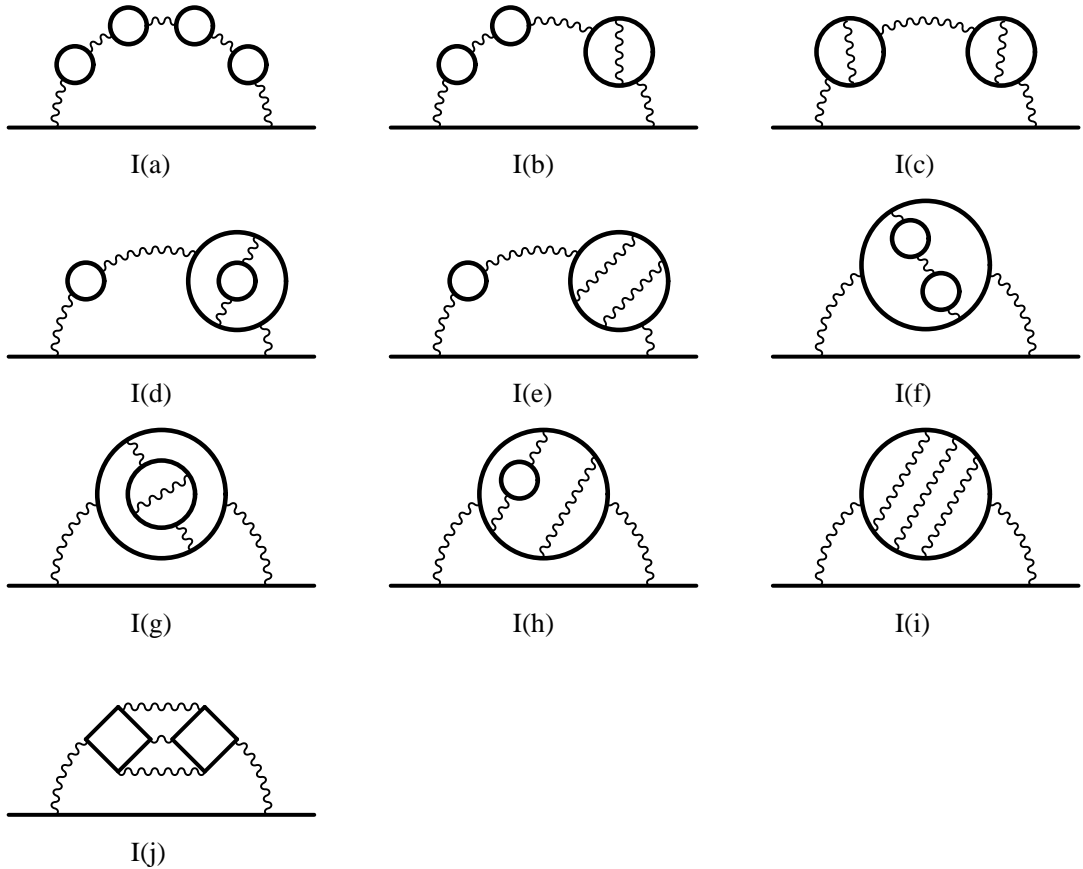}  
\caption{Set I. 
There are 208 Feynman diagrams in this set.
\label{fig:set1}
}
\end{figure} 

Set I consists of 208 diagrams of the form shown 
in Fig.~\ref{fig:set1}. 
Each of these diagrams is obtained by inserting 
vacuum-polarization or light-by-light-scattering subdiagrams 
into a q-type diagram of the second order. 
Set I can be classified into ten gauge-invariant subsets. 
\begin{description} 
\item[I(a)] 
A diagram contains four vacuum-polarizations of the second order. 
\item[I(b)] 
Each diagram contains 
a fourth-order vacuum-polarization 
and two vacuum-polarizations of the second order. 
\item[I(c)] 
Each diagram contains 
two vacuum-polarizations of the fourth order. 
\item[I(d)] 
Each diagram contains 
a second-order vacuum-polarization 
and a sixth-order vacuum-polarization which consists 
of two lepton loops. 
\item[I(e)] 
Each diagram contains 
a second-order vacuum-polarization 
and a sixth-order vacuum-polarization which consists 
of a single lepton loop. 
\item[I(f)] 
Each diagram contains 
an eighth-order vacuum-polarization which consists 
of three lepton loops. 
\item[I(g)] 
Each diagram contains 
an eighth-order vacuum-polarization which contains 
a fourth-order vacuum-polarization as its subdiagram. 
\item[I(h)]
Each diagram contains 
an eighth-order vacuum-polarization which contains 
a second-order vacuum polarization as its subdiagram. 
\item[I(i)] 
Each diagram contains 
an eighth-order vacuum-polarization which consists 
of a single lepton loop. 
\item[I(j)] 
Each diagram contains 
an eighth-order vacuum-polarization which consists 
of two light-by-light-scattering subdiagrams. 
\end{description} 
\begin{figure}[htb] 
\includegraphics[scale=0.8]{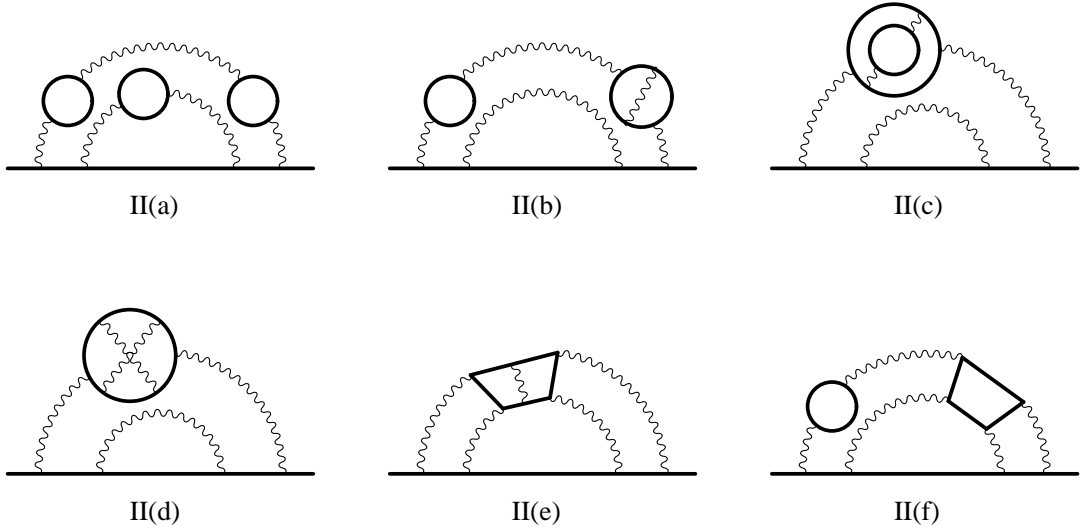} 
\caption{Set II. 
There are 600 Feynman diagrams in this set.
\label{fig:set2}
}
\end{figure} 

Set II consists of 600 diagrams of the form shown 
in Fig.~\ref{fig:set2}, 
each of which is obtained by inserting vacuum-polarization 
subdiagrams and/or a light-by-light-scattering subdiagram 
into a q-type diagram of the fourth order. 
Set II can be further classified into six gauge-invariant subsets.  
\begin{description} 
\item[II(a)] 
Each diagram contains 
three vacuum-polarizations of the second order. 
\item[II(b)] 
Each diagram contains 
a second-order vacuum-polarization 
and a fourth-order vacuum-polarization. 
\item[II(c)] 
Each diagram contains 
a sixth-order vacuum-polarization which contains an internal lepton loop. 
\item[II(d)] 
Each diagram contains 
a sixth-order vacuum-polarization which consists of only one lepton loop. 
\item[II(e)] 
Each diagram contains 
a sixth-order light-by-light-scattering subdiagram. 
\item[II(f)] 
Each diagram contains 
a fourth-order light-by-light-scattering subdiagram 
and a second-order vacuum-polarization. 
\end{description} 
\begin{figure}[htb] 
\includegraphics[scale=0.8]{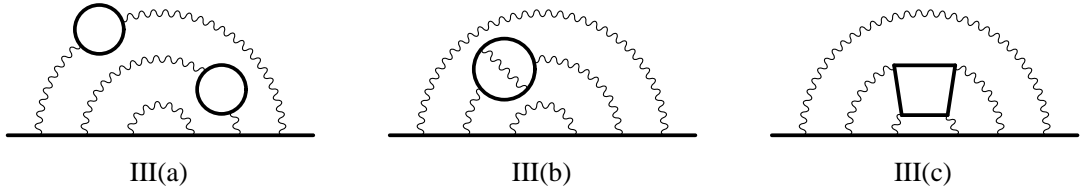}  
\caption{Set III. 
There are 1140 diagrams in this set.
\label{fig:set3} 
}
\end{figure} 

Set III consists of 1140 diagrams of the form shown 
in Fig.~\ref{fig:set3}, 
each of which is obtained by inserting a vacuum-polarization 
subdiagram and/or a light-by-light-scattering subdiagram 
into a sixth-order q-type diagram. 
Set III can be further classified into three gauge invariant subsets. 
\begin{description} 
\item[III(a)] 
Each diagram contains 
two vacuum-polarizations of the second order. 
\item[III(b)] 
Each diagram contains 
a fourth-order vacuum-polarization. 
\item[III(c)] 
Each diagram contains 
a fourth-order light-by-light-scattering subdiagram. 
\end{description} 

\begin{figure}[htb] 
\includegraphics[scale=0.8]{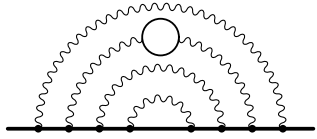}  
\caption{Set IV. 
There are 2072 Feynman diagrams in this set.
\label{fig:set4}
}
\end{figure} 

Set IV consists of 2072 diagrams of the form shown 
in Fig.~\ref{fig:set4}, 
each of which is obtained by inserting 
a second-order vacuum-polarization 
into an eighth-order q-type diagram. 

\begin{figure}[htb] 
\includegraphics[scale=0.8]{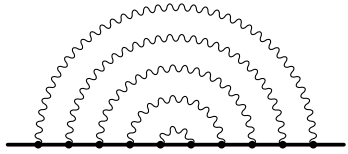}  
\caption{Set V. 
There are 6354 Feynman diagrams in this set.
\label{fig:set5}
}
\end{figure}

Set V in Fig.~\ref{fig:set5} consists of 6354 vertex diagrams 
of the q-type of the tenth order. 

\begin{figure}[htb] 
\includegraphics[scale=0.8]{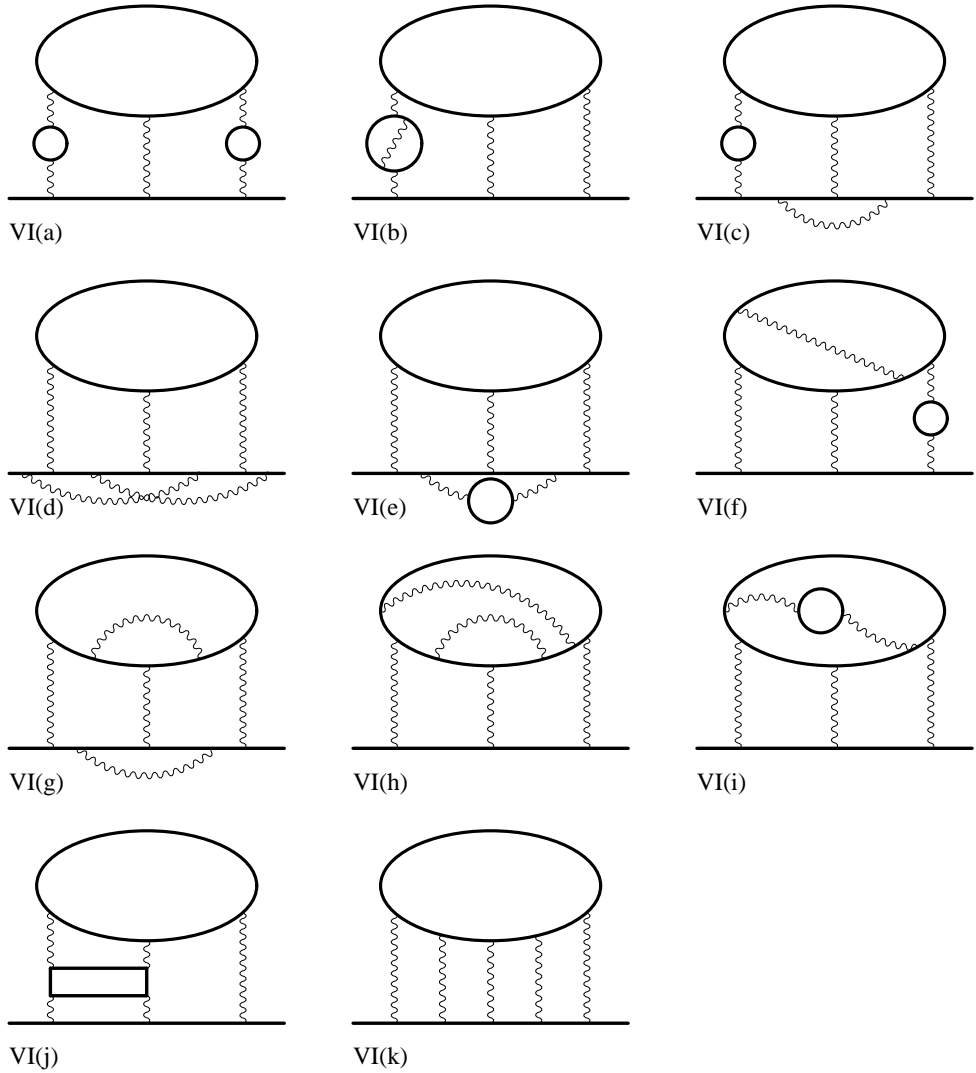}  
\caption{Set VI. 
There are 2298 Feynman diagrams in this set.
\label{fig:set6}
}
\end{figure} 

Set VI consists of 2298 Feynman diagrams of the form shown 
in Fig.~\ref{fig:set6}. 
They contain a light-by-light-scattering subdiagram, one of 
whose photon lines is supposed to be external. 
We call them as \textit{l-by-l}-type diagrams hereafter. 
This set is further classified into eleven gauge invariant subsets. 
Each subset of diagrams also includes radiative corrections of 
respective types except for the subset VI(k). 
\begin{description} 
\item[VI(a)] 
Each diagram is obtained by inserting 
two vacuum-polarizations of the second order 
into a l-by-l-type diagram. 
\item[VI(b)] 
Each diagram is obtained by inserting 
a fourth-order vacuum-polarization
into a l-by-l-type diagram. 
\item[VI(c)] 
Each diagram is obtained by inserting a second-order 
vacuum-polarization into one of virtual photon lines coming 
out of the lepton loop of the l-by-l-type diagram 
and attaching one internal photon line to the open lepton line. 
\item[VI(d)] 
Each diagram is obtained by attaching 
two internal photon lines 
to the open lepton line of a l-by-l-type diagram.
\item[VI(e)]  
Each diagram is obtained by attaching 
an internal photon line with a second-order vacuum-polarization inserted 
to the open lepton line of a l-by-l-type diagram. 
\item[VI(f)] 
Each diagram is obtained by first attaching 
an internal photon line within the lepton loop 
of a l-by-l-type diagram 
and then inserting a second-order vacuum-polarization 
into one of the three photon lines 
which connect the lepton loop and the open lepton line. 
\item[VI(g)] 
Each diagram is obtained by first attaching 
an internal photon line within the lepton loop 
of a l-by-l-type diagram 
and then attaching an internal photon line to the open lepton line. 
\item[VI(h)] 
Each diagram is obtained by attaching 
two internal photon lines within the lepton loop 
of a l-by-l-type diagram. 
\item[VI(i)] 
Each diagram is obtained by attaching 
an internal photon line with a second-order vacuum-polarization inserted 
to the lepton loop of a l-by-l-type diagram. 
\item[VI(j)] 
Each diagram is obtained by inserting 
a light-by-light-scattering subdiagram of the fourth order 
into a l-by-l-type diagram 
\item[VI(k)] 
Each diagram contains a light-by-light-scattering amplitude 
with six photon legs. 
\end{description}  

%================================================================
%================================================================
\section{Alternate formulae for $U$ and $B$ polynomials} 
%================================================================
\label{sec:detail}

Definitions (\ref{eq:DefOfU}) and (\ref{eq:DefOfB}) 
are useful for the programs like MATHEMATICA and MAPLE. 
However, they turned out to be clumsy for developing programs 
by languages such as C++ and FORTRAN. 
This is why it is useful to find alternate formulae for 
$U$ and $B$ polynomials. 
Our strategy is to directly provide compact expression for each 
term which is homogeneous polynomial of Feynman parameters $\{z_i\}$. 

%----------------------------------------------------------------
\subsection{Concise formula for $U$}
\label{sec:detail:U}

Let us first consider the $U$ polynomial. 
We introduce the circuit matrix characterized 
by the chain indices $\alpha$ 
instead of the line indices $j$ by 
\begin{equation} 
	\xi_{\alpha,\,r} = \xi_{j,\,r} 
	\quad 
	(j \in \alpha) \, . 
\label{eq:circuitMatrixForChain}
\end{equation} 
This is unambiguously defined in our convention for orientations 
of chains and lines. 
Then, Eq.~(\ref{eq:DefOfUst}) can be rewritten in the form which 
expresses that $U$ and $U_{rs}$ are determined 
solely by the structure of chains: 
\begin{equation} 
	U_{rs} 
	= 
	\sum_{{\rm all}\ \alpha} \xi_{\alpha,\,r}\,w_\alpha\,\xi_{\alpha,\,s} 
	\,, 
\label{eq:U_cmByChain}
\end{equation} 
where $w_\alpha$ is defined by $w_\alpha = \sum_{i\in\alpha} z_i$. 
Now, $U$ of Eq.~(\ref{eq:DefOfU}) can be expanded as 
\begin{eqnarray} 
	U &=& \det_{1 \le r,\,s \le n}\,U_{rs} 
\nonumber \\ 
	&=& 
	\sum_{\alpha_1} w_{\alpha_1}\, \xi_{\alpha_1,\,1} 
	\cdots 
	\sum_{\alpha_{n}} w_{\alpha_{n}}\, \xi_{\alpha_{n},\,n} 
\nonumber \\ 
	&& \quad 
	\times 
	\sum_{\sigma \in {\mathfrak S}_{n}} 
	\varepsilon(\sigma)\, 
	\xi_{\alpha_1,\,\sigma(1)} \cdots  \xi_{\alpha_{n},\,\sigma(n)}
	\,, 
\label{eq:calculation_for_U} 
\end{eqnarray} 
where ${\mathfrak S}_{n}$ denotes the permutation group of 
degree $n$ and $\varepsilon(\sigma) = \pm 1$ is the signature of 
$\sigma \in {\mathfrak S}_{n}$. 
The right-hand side of the last equality shows that the terms with 
$\alpha_r = \alpha_s$ for $r \ne s$ vanishes. 
Thus $U$ is a homogeneous polynomial of degree $n$, 
where each monomial can be at most linear with respect to each 
$w_\alpha$. 
Such a monomial is characterized by a combination 
$\left\{\alpha_1,\,\cdots,\,\alpha_{n}\right\}$ 
whose elements are picked up from the sets 
$\left\{1,\,\cdots,\,3 n - 3\right\}$. 
They can be ordered as $\alpha_1 < \cdots < \alpha_{n}$ 
taking the permutation of the indices attached to the circuits 
in Eq.~(\ref{eq:calculation_for_U}) into account. 
Then, Eq.~(\ref{eq:calculation_for_U}) becomes 
\begin{eqnarray} 
	U &=& 
	\sum_{1 \le \alpha_1 < \cdots < \alpha_{n} \le 3 n - 3}  
	w_{\alpha_1} \cdots w_{\alpha_{n}}  
\nonumber \\ 
	&& 
\qquad \qquad \qquad 
	\times 
	\sum_{\sigma^\prime \in {\mathfrak S}_{n}}  
	\xi_{\alpha_1,\,\sigma^\prime(1)}  
	\cdots 
	\xi_{\alpha_{n},\,\sigma^\prime(n)}  
\nonumber \\ 
	&& 
\qquad \qquad \qquad \qquad 
	\times 
	\sum_{\sigma \in {\mathfrak S}_{n}} 
	\varepsilon(\sigma)\,  
	\xi_{\alpha_1,\,\sigma(\sigma^\prime(1))} 
	\cdots 
	\xi_{\alpha_{n},\,\sigma(\sigma^\prime(n))} 
	 \,. 
\label{eq:calculation_for_U_2} 
\end{eqnarray} 
We replace the sum over $\sigma \in {\mathfrak S}_{n}$ by the sum over 
$\sigma^{\prime\prime} \equiv \sigma \circ \sigma^\prime \in {\mathfrak S}_{n}$. 
Then, noting 
$\varepsilon(\sigma^{\prime\prime})=\varepsilon(\sigma^\prime)\,\varepsilon(\sigma)$, 
the quantities appearing in the second line and the third line on 
the right-hand side of Eq.~(\ref{eq:calculation_for_U_2}) turn out 
to factorize and coincide with each other. 
Thus, we get the following formula for $U$: 
\begin{equation} 
	U 
	= 
	\sum_{1 \le \alpha_1 < \cdots < \alpha_{n} \le 3 n - 3}  
	w_{\alpha_1} \cdots w_{\alpha_{n}} \, 
	\left(A(\alpha_1,\,\cdots,\,\alpha_{n})\right)^2 \, ,  
\label{eq:new_formula_for_U} 
\end{equation} 
with 
\begin{equation} 
	A(\alpha_1,\,\cdots,\,\alpha_{n}) 
	\equiv 
	\sum_{\sigma \in {\mathfrak S}_{n}} 
	\varepsilon(\sigma)\,  
	\xi_{\alpha_1,\,\sigma(1)} 
	\cdots 
	\xi_{\alpha_{n},\,\sigma(n)} \, . 
\label{eq:def_of_quantity_A} 
\end{equation}

%----------------------------------------------------------------
\subsection{Concise formula for $B_{\alpha\beta}$}
\label{sec:detail:Bij}

We also find a convenient formula for $B$ polynomials following 
a similar manipulation as above. 
We start with the expression 
\begin{equation} 
	B_{\alpha \beta} 
	= 
	U \sum_{r,\,s = 1}^{n} 
	\xi_{\alpha,\,r} \left(U^{-1}\right)_{rs}\,\xi_{\beta,\,s} \, , 
\label{eq:BbyChain} 
\end{equation} 
showing that they are also determined solely 
by the associated chain diagram. 
By deferring its derivation below, we obtain a compact formula 
\begin{eqnarray} 
	B_{\alpha,\,\beta} 
	&=& 
	\sum_{\alpha_1 < \cdots < \alpha_{n - 1}} 
	w_{\alpha_1} \cdots w_{\alpha_{n - 1}} 
\nonumber \\ 
	&& 
\qquad \qquad \qquad 
	\times 
	A(\alpha,\,\alpha_1,\,\cdots,\,\alpha_{n - 1})\,  
	A(\beta,\,\alpha_1,\,\cdots,\,\alpha_{n - 1})\, ,  
\label{eq:new_formular_for_Bfunction}
\end{eqnarray} 
with 
$A(\alpha,\,\alpha_1,\,\cdots,\,\alpha_{n - 1})$ 
given in Eq.~(\ref{eq:def_of_quantity_A}). 
For $\alpha = \beta$, there is a particular equation which follows 
from Eqs.~(\ref{eq:new_formula_for_U}) 
and (\ref{eq:new_formular_for_Bfunction}) 
\cite{Cvitanovic:1974uf}, 
\begin{equation} 
	B_{\alpha\,\alpha} = \frac{\del U}{\del w_\alpha} \, . 
\label{eq:Baa}
\end{equation} 
Previously most hand calculation has been done using 
the Nakanishi formula 
\begin{equation} 
	B_{\alpha\,\beta} 
	= 
	\sum_{c} \xi_{\alpha,\,c}\,\xi_{\beta,\,c}\,U_{{\cal G} /c} \, , 
\label{eq:NFforB} 
\end{equation} 
where the sum runs over all possible circuits 
(not necessarily limited to those of the fundamental set of circuits), 
$\xi_{\alpha,\,c}=(1,-1,0)$ is a projector to whether the chain 
$\alpha$ runs (along, against, outside of) the loop $c$, 
and $U_{{\cal G} /c}$ is the $U$ polynomial of the reduced diagram 
${\cal G} /c$ obtained by shrinking the circuit $c$ and the vertices 
on it to a single vertex. 
The formula (\ref{eq:NFforB}) requires a topological manipulation 
to pick up all circuits in a graph, 
while the formula (\ref{eq:new_formular_for_Bfunction}) 
enables us to calculate $B$ polynomials algebraically.

%----------------------------------------------------------------
\vskip 2ex
\noindent
\textbf{Derivation of Eq.~(\ref{eq:new_formular_for_Bfunction}).}
%----------------------------------------------------------------

We derive the formula (\ref{eq:new_formular_for_Bfunction}) 
for $B_{\alpha\,\beta}$. 
For that purpose, we introduce a set of functions 
$\{f_r\}_{r = 1, \cdots, n}$ on the domain $\{1,\,\cdots,n - 1\}$ by 
\begin{equation} 
	f_r(x) 
	= 
	\begin{cases}
	 x     & \qquad (1 \le x \le r - 1) \\ 
	 x + 1 & \qquad (r \le x \le n - 1) \\ 
	\end{cases} \,.
\end{equation} 
Then, the elements of the $(n - 1) \times (n - 1)$ minor matrix 
$U_{\widehat{rs}}$ obtained by eliminating the $s$th column and 
the $r$th raw from $\left\{U_{sr}\right\}_{s,\,r = 1,\,\cdots,\,n}$ 
are given by (also using the symmetric property of $U_{rs}$) 
\begin{equation} 
	\left( U_{\widehat{rs}} \right)_{xy} 
	= 
	U_{f_r(x),\,f_s(y)} \, . 
\label{eq:expr_for_minor}
\end{equation} 
Inserting Eq.~(\ref{eq:expr_for_minor}) into 
\begin{equation} 
	U\,\left(U^{-1}\right)_{rs} 
	= 
	(-1)^{r + s}\, 
	\det_{1 \le x,\,y \le (n - 1)} 
	\left((U_{\widehat{r s}})_{x y}\right) \, ,  
\label{eq:UInvTimesU}
\end{equation} 
the part $U (U^{-1})_{rs}$ in Eq.~(\ref{eq:BbyChain}) can be written as 
\begin{eqnarray} 
	U (U^{-1})_{rs} 
	&=& 
	(-1)^{r + s} 
	\sum_{\sigma \in {\mathfrak S}_{n - 1}} 
	\varepsilon(\sigma)\, 
	U_{f_r(1),\,f_s(\sigma(1))} \cdots 
	U_{f_r(n - 1),\,f_s(\sigma(n - 1))}  
\nonumber \\ 
	&=& 
	(-1)^{r + s} 
	\sum_{\alpha_1} w_{\alpha_1}\, \xi_{\alpha_1,\,f_r(1)}  
	\cdots 
	\sum_{\alpha_{n - 1}} 
	w_{\alpha_{n - 1}}\, \xi_{\alpha_{n - 1},\,f_r(n - 1)} 
\nonumber \\ 
	&& 
\qquad 
	\times 
	\sum_{\sigma \in {\mathfrak S}_{n - 1}} 
	\varepsilon(\sigma)\, 
	\xi_{\alpha_1,\,f_s(\sigma(1))} \cdots 
	\xi_{\alpha_{n - 1},\,f_s(\sigma(n - 1))} 
\nonumber \\ 
	&=& 
	(-1)^{r + s} 
	\sum_{\alpha_{r^\prime} \ne \alpha_{s^\prime}}  
	w_{\alpha_1} \cdots w_{\alpha_{n - 1}}\, 
	\xi_{\alpha_1,\,f_r(1)} \cdots \xi_{\alpha_{n - 1},\,f_r(n - 1)} 
\nonumber \\ 
	&& 
\qquad \quad 
	\times 
	\sum_{\sigma \in {\mathfrak S}_{n - 1}} 
	\varepsilon(\sigma)\, 
	\xi_{\alpha_1,\,f_s(\sigma(1))} \cdots 
	\xi_{\alpha_{n - 1},\,f_s(\sigma(n - 1))} 
\nonumber \\ 
	&=& 
	(-1)^{r + s} 
	\sum_{\alpha_1 < \cdots < \alpha_{n - 1}} 
	w_{\alpha_1} \cdots w_{\alpha_{n - 1}} 
\nonumber \\ 
	&& 
\qquad \quad 
	\times 
	\sum_{\sigma^\prime \in {\mathfrak S}_{n - 1}} 
	\xi_{\alpha_{\sigma^\prime(1)},\,f_r(1)} \cdots 
	\xi_{\alpha_{\sigma^\prime(n - 1)},\,f_r(n - 1)} 
\nonumber \\ 
	&& 
\qquad \qquad \quad 
	\times 
	\sum_{\sigma \in {\mathfrak S}_{n - 1}} 
	\varepsilon(\sigma)\, 
	\xi_{\alpha_{\sigma^\prime(1)},\,f_s(\sigma(1))} \cdots 
	\xi_{\alpha_{\sigma^\prime(n - 1)},\,f_s(\sigma(n - 1))} 
\nonumber \\ 
	&=& 
	(-1)^{r + s} 
	\sum_{\alpha_1 < \cdots < \alpha_{n - 1}} 
	w_{\alpha_1} \cdots w_{\alpha_{n - 1}} 
\nonumber \\ 
	&& 
\qquad \quad 
	\times 
	\sum_{\sigma^\prime \in {\mathfrak S}_{n - 1}} 
	\xi_{\alpha_1,\,f_r(\sigma^\prime(1))} \cdots 
	\xi_{\alpha_{n - 1},\,f_r(\sigma^\prime(n - 1))} 
\nonumber \\ 
	&& 
\qquad \qquad \quad 
	\times 
	\sum_{\sigma \in {\mathfrak S}_{n - 1}} 
	\varepsilon(\sigma)\, 
	\xi_{\alpha_1,\,f_s((\sigma \circ \sigma^\prime)(1))} \cdots 
	\xi_{\alpha_{n - 1},\,f_s((\sigma \circ \sigma^\prime)(n - 1))} 
\nonumber \\ 
	&=& 
	(-1)^{r + s} 
	\sum_{\alpha_1 < \cdots < \alpha_{n - 1}} 
	w_{\alpha_1} \cdots w_{\alpha_{n - 1}} 
\nonumber \\ 
	&& 
\qquad \quad 
	\times 
	\left( 
		\sum_{\sigma \in {\mathfrak S}_{n - 1}} 
		\varepsilon(\sigma)\, 
		\xi_{\alpha_1,\,f_r(\sigma(1))} \cdots 
		\xi_{\alpha_{n - 1},\,f_r(\sigma(n - 1))} 
	\right) 
\nonumber \\ 
	&& 
\qquad \quad 
	\times 
	\left( 
		\sum_{\sigma \in {\mathfrak S}_{n - 1}} 
		\varepsilon(\sigma)\, 
		\xi_{\alpha_1,\,f_s(\sigma(1))} \cdots 
		\xi_{\alpha_{n - 1},\,f_s(\sigma(n - 1))} 
	\right) \, .  
\label{eq:eval_cofactor_matrix} 
\end{eqnarray} 
By inserting this expression into Eq.~(\ref{eq:BbyChain}), we get 
\begin{equation} 
	B_{\alpha\,\beta} 
	= 
	\sum_{\alpha_1 < \cdots < \alpha_{n - 1}} 
	w_{\alpha_1} \cdots w_{\alpha_{n - 1}}\, 
	B(\alpha;\,\alpha_1,\,\cdots,\,\alpha_{n - 1})\, 
	B(\beta;\,\alpha_1,\,\cdots,\,\alpha_{n - 1}) \, , 
\end{equation} 
where 
\begin{equation} 
	B(\alpha;\,\alpha_1,\,\cdots,\,\alpha_{n - 1}) 
	\equiv 
	\sum_{r = 1}^{n} 
	(-1)^{r + 1}\,\xi_{\alpha,\,r} 
	\sum_{\sigma \in {\mathfrak S}_{n - 1}} 
	\varepsilon(\sigma)\, 
	\xi_{\alpha_1,\,f_r(\sigma(1))} \cdots 
	\xi_{\alpha_{n - 1},\,f_r(\sigma(n - 1))} \, . 
\end{equation} 
The remained task is to demonstrate that 
this $ B(\alpha;\,\alpha_1,\,\cdots,\,\alpha_{n - 1})$ 
coincides with $A(\alpha,\,\alpha_1\,\cdots,\,\alpha_{n - 1})$ 
appearing in the coefficient (\ref{eq:def_of_quantity_A}) 
of each monomial of $U$. 
 For that purpose, the summation over all permutations 
of the circuits in a fundamental set 
is replaced by that over all permutations 
of indices distinguishing the chains; 
\begin{eqnarray} 
	B(\alpha;\,\alpha_1,\,\cdots,\,\alpha_{n - 1}) 
	&=& 
	\sum_{r = 1}^{n} 
	(-1)^{r + 1} \xi_{\alpha,\,r} 
	\sum_{\sigma \in {\mathfrak S}_{n - 1}} 
	\varepsilon(\sigma)\, 
	\xi_{\alpha_{\sigma(1)},\,f_r(1)} \cdots 
	\xi_{\alpha_{\sigma(n - 1)},\,f_r(n - 1)} 
\nonumber \\ 
	&=& 
	\sum_{r = 1}^{n} 
	\sum_{\sigma \in {\mathfrak S}_{n - 1}} 
	(-1)^{r + 1} \varepsilon(\sigma)\, 
	\xi_{\alpha_{\sigma(1)},\,1} \cdots 
	\xi_{\alpha_{\sigma(r - 1)},\,r - 1}\,  
	\xi_{\alpha,\,r} 
\nonumber \\ 
	&& 
\qquad \qquad \qquad \qquad 
	\times 
	\xi_{\alpha_{\sigma(r)},\,r + 1} \cdots 
	\xi_{\alpha_{\sigma(n - 1),\,n}} 
\nonumber \\ 
	&=& 
	\sum_{\sigma \in {\mathfrak S}_{n}} 
	\varepsilon(\sigma)\,
	\xi_{\alpha,\,\sigma(1)}\, \xi_{\alpha_1,\,\sigma(2)} 
	\cdots 
	\xi_{\alpha_{n - 1},\,\sigma(n)} 
\nonumber \\ 
	&=& 
	A(\alpha,\,\alpha_1,\,\cdots,\,\alpha_{n - 1}) \, . 
\label{eq:B_equalTo_A} 
\end{eqnarray} 
In the above, we use the fact that a sequence of permutations 
\begin{eqnarray} 
	&& 
	\left[ 
	\alpha_{\sigma(1)},\,\cdots,\,\alpha_{\sigma(r - 1)},\, 
	\alpha,\,\alpha_{\sigma(r)},\,\cdots,\,\alpha_{\sigma(n - 1)} 
	\right] 
\nonumber \\ 
	&& 
\qquad 
	\mapsto  
	\left[ 
	\alpha,\, 
	\alpha_{\sigma(1)},\,\cdots,\,\alpha_{\sigma(r - 1)},\, 
	\alpha_{\sigma(r)},\,\cdots,\,\alpha_{\sigma(n - 1)} 
	\right] 
\nonumber \\ 
	&& 
\qquad 
	\mapsto 
	\left[\alpha,\,\alpha_1,\,\cdots,\,\alpha_{n - 1}\right] 
	\, , 
\end{eqnarray} 
for $\sigma \in {\mathfrak S}_{n - 1}$, combines to form all 
possible permutations of degree $n$ and each step produces 
a signature $(-1)^{r + 1}$ and $\varepsilon(\sigma)$ respectively. 
Therefore, we obtain the desired result 
(\ref{eq:new_formular_for_Bfunction}).

%----------------------------------------------------------------
\subsection{Concise formula for $C_{ij}$}
\label{sec:detail:Cij}

A $C$-polynomial is a function characterized by a pair 
$(j_1,\,j_2)$ of two indices of lepton lines 
\begin{align}
	C_{j_1\,j_2}
	& \equiv \frac{1}{U} \widetilde{C}_{j_1\,j_2}, 
\label{eq:defCprimePoly} \\
	\widetilde{C}_{j_1\,j_2} 
	& = 
	\frac{1}{U} 
	{\sum_{k_1 < k_2}}^\prime z_{k_1}\,z_{k_2} 
	\left( 
		B^\prime_{k_1\,j_1}\,B^\prime_{k_2\,j_2} 
		- 
		B^\prime_{k_1\,j_2}\,B^\prime_{k_2\,j_1} 
	\right) \,, 
\label{eq:defCPoly}
\end{align}
where the summation ranges over 
all pairs $(k_1,\,k_2)$ of the indices of lepton lines. 
The reason why we call this function as a polynomial will 
be clarified shortly. 
From its definition, $\widetilde{C}_{j_1 j_2} = -\widetilde{C}_{j_2 j_1}$. 
Thus, we can assume that $j_1 < j_2$ without loss of generality. 

The expression (\ref{eq:defCPoly}) of $\widetilde{C}_{j_1 j_2}$ 
needs a little bit care. 
The factor $\dfrac{1}{U}$ prior to the summation   
might imply that $\widetilde{C}_{j_1\,j_2}$ develops a singularity 
like $\dfrac{1}{U}$ when $U \to 0$. 
If it were the case, 
the maximally contracted terms containing 
$\widetilde{C}_{j_1 j_2}$ 
would dominate over the other terms in the UV-limit. 
This should not be the case 
since the degree of ultraviolet singularities 
of the terms containing $C$-polynomials,  
would be higher than that expected 
from the original expression of Feynman amplitude. 
A closer examination shows that 
one factor of $U$ factorizes out from 
$ \left( 
	B^\prime_{k_1\,j_1}\,B^\prime_{k_2\,j_2} 
	- 
	B^\prime_{k_1\,j_2}\,B^\prime_{k_2\,j_1} 
 \right)$ 
in Eq.~(\ref{eq:defCPoly}). 
Thus, $\widetilde{C}_{j_1\,j_2}$ are actually polynomials, 
and the maximally contracted terms of $C$-polynomials 
have the same degree of singularity 
as that of the other maximally contracted terms. 
The factorization of $U$ could be realized 
at the numerical level with use of Eq.~(\ref{eq:defCPoly}). 
However, to avoid round-off errors, 
it is desirable to obtain an analytic expression 
that calculates $\widetilde{C}_{j_1 j_2}$ as a polynomial. 
Below, 
we write down such an expression that 
also enables us to control the UV limit of the $C$-polynomials. 

Since $B^\prime_{k\,j}$ differs from $B_{k\,j}$ when $j = k$ 
(See Eq.~(\ref{eq:DefBprime})), 
we define the quantity 
\begin{equation} 
	c(\,k_1,\,k_2;\,j_1,\,j_2) 
	\equiv 
	\frac{1}{U}\, 
	z_{k_1}\,z_{k_2} 
	\left( 
		B^\prime_{k_1,\,j_1}\,B^\prime_{k_2,\,j_2} 
		-
		B^\prime_{k_1,\,j_2}\,B^\prime_{k_2,\,j_1} 
	\right) \, . 
\label{eq:cij:defc}
\end{equation} 
and distinguish the six cases; 
\begin{description}
\item[\rm I\,:\ $j_1 = k_1 < j_2 = k_2$]
\[
	c(j_1,\,j_2;\,j_1,\,j_2) 
	= 
	z_{j_1}\,z_{j_2}\, 
	B_{\alpha_{j_1},\,\alpha_{j_1};\,\alpha_{j_2},\,\alpha_{j_2}} 
	- 
	\left( 
		z_{j_1}\,B_{\alpha_{j_1},\,\alpha_{j_1}}  
		+ z_{j_2}\,B_{\alpha_{j_2},\,\alpha_{j_2}}  
	\right) 
	+ 
	U \, , 
\]
\item[\rm IIa\,:\ $j_1 \ne k_1 < j_2 = k_2$]
\[
	c(k_1,\,j_2;\,j_1,\,j_2) 
	= 
	z_{k_1}\,z_{j_2}\, 
	B_{\alpha_{k_1},\,\alpha_{j_1};\,\alpha_{j_2},\,\alpha_{j_2}} 
	- 
	z_{k_1}\,B_{\alpha_{k_1},\,\alpha_{j_1}} \, , 
\]
\item[\rm IIb\,:\ $j_1 = k_1 < j_2 \ne k_2$]
\[
	c(j_1,\,k_2;\,j_1,\,j_2) 
	= 
	z_{j_1}\,z_{k_2}\, 
	B_{\alpha_{j_1},\,\alpha_{j_1};\,\alpha_{k_2},\,\alpha_{j_2}} 
	- 
	z_{k_2}\,B_{\alpha_{k_2},\,\alpha_{j_2}} \, , 
\]
\item[\rm IIc\,:\ $j_1 < j_2 = k_1 < k_2$]
\[
	c(j_2,\,k_2;\,j_1,\,j_2) 
	= 
	- 
	z_{j_2}\,z_{k_2}\, 
%  B_{\alpha_{j_2},\,\alpha_{j_1};\,\alpha_{k_2},\,\alpha_{j_2}}
	B_{\alpha_{j_2},\,\alpha_{j_2};\,\alpha_{k_2},\,\alpha_{j_1}}  
	+ 
	z_{k_2}\,B_{\alpha_{k_2},\,\alpha_{j_1}} \, , 
\]
\item[\rm IId\,:\ $k_1 < k_2 = j_1 < j_2$]
\[
	c(k_1,\,j_1;\,j_1,\,j_2) 
	= 
	- 
	z_{j_1}\,z_{k_1}\, 
%  B_{\alpha_{k_1},\,\alpha_{j_1};\,\alpha_{j_1},\,\alpha_{j_2}} 
	B_{\alpha_{j_1},\,\alpha_{j_1};\,\alpha_{k_1},\,\alpha_{j_2}} 
	+ 
	z_{k_1}\,B_{\alpha_{k_1},\,\alpha_{j_2}} \, , 
\]
\item[\rm III\,:\ all\ $j_1,\,j_2,\,k_1,\,k_2$\ are\ different]
\[
	c(k_1,\,k_2;\,j_1,\,j_2) 
	= 
	z_{k_1}\,z_{k_2}\, 
	B_{\alpha_{k_1},\,\alpha_{j_1};\,\alpha_{k_2},\,\alpha_{j_2}} \, , 
\]
\label{eq:c_6cases}
\end{description}
where 
\begin{equation} 
	B_{\alpha,\,\beta;\,\alpha^\prime,\,\beta^\prime} 
	\equiv 
	\frac{1}{U} 
	\left( 
		B_{\alpha,\,\beta}\, B_{\alpha^\prime,\,\beta^\prime} 
		- 
		B_{\alpha,\,\beta^\prime}\, B_{\alpha^\prime,\,\beta} 
	\right) \, , 
\label{eq:DefOfB4Tensor}
\end{equation} 
and each $\alpha_j$ denotes the index of the chain containing 
the lepton line $l_j$. 

The remained task is to find out 
a convenient expression of 
$B_{\alpha,\,\beta;\,\alpha^\prime,\,\beta^\prime}$ 
as a polynomial of $w_\alpha$. 
 Expressing $U^{-1}\,U_{rs}$ in Eq.~(\ref{eq:DefOfB4Tensor}) by 
Eq.~(\ref{eq:UInvTimesU}), 
$B_{\alpha,\,\beta;\,\alpha^\prime,\,\beta^\prime}$ 
reduces to a polynomial 
\begin{eqnarray} 
	B_{\alpha,\,\beta;\,\alpha^\prime,\,\beta^\prime} 
	&=& 
	\sum_{1 \le r < s \le n} 
	\sum_{1 \le r^\prime < s^\prime \le n} 
	\left( 
		\xi_{\alpha r} \xi_{\alpha^\prime s} 
		- \xi_{\alpha s} \xi_{\alpha^\prime r} 
	\right) 
	\left( 
		\xi_{\beta r^\prime} \xi_{\beta^\prime s^\prime} 
		- \xi_{\beta s^\prime} \xi_{\beta^\prime r^\prime} 
	\right) 
\nonumber \\ 
	&& 
\qquad \qquad \qquad 
	\times 
	(-1)^{r + r^\prime + s + s^\prime} 
	\det \left(U(rs|r^\prime s^\prime)\right) \, . 
\label{eq:B4_ito_det}
\end{eqnarray} 
Here $\det \left(U(rs|r^\prime s^\prime)\right)$ 
is the determinant of the $(n - 2) \times (n - 2)$ 
matrix $\left\{U(rs|r^\prime s^\prime)_{xy}\right\}$ 
obtained from 
$\left\{U_{r^{\prime\prime} s^{\prime\prime}}\right\}_{1 \le r^{\prime\prime},\,s^{\prime\prime} \le n}$ 
by eliminating the $r$th and $s$th columns 
and the $r^\prime$th and $s^\prime$th raws. 
If we define the function $f_{r < s}(x)$ on the domain 
$\left\{1,\,\cdots,\,n - 2\right\}$ by 
\begin{equation} 
	f_{r<s}(x) 
	= 
	\begin{cases}
	x     & \qquad 1 \le x \le r - 1 ,\\ 
	x + 1 & \qquad r \le x \le s - 2 ,\\ 
	x + 2 & \qquad s - 1 \le x \le n - 2 ,
	\end{cases}
\end{equation} 
the matrix elements $U(rs|r^\prime s^\prime)_{xy}$ are expressed  
in terms of $U_{r^{\prime\prime} s^{\prime\prime}}$ as 
\begin{equation} 
	U(rs|r^\prime s^\prime)_{xy} 
	= 
	U_{f_{r<s}(x),\,f_{r^\prime < s^\prime}(y)} \, . 
\end{equation} 
The application of the similar manipulation 
as that gives Eq.~(\ref{eq:eval_cofactor_matrix}) 
to $\det \left(U(rs|r^\prime s^\prime)\right)$ yields 
\begin{eqnarray} 
	\det \left(U(rs|r^\prime s^\prime)\right) 
	&=& 
	\sum_{\alpha_1 < \cdots < \alpha_{n - 2}} 
	w_{\alpha_1} \cdots w_{\alpha_{n - 2}} 
\nonumber \\ 
	&& \quad 
	\times 
	\left( 
	\sum_{\sigma \in {\mathfrak S}_{n - 2}} 
	\varepsilon(\sigma)\,  
	\xi_{\alpha_1,\,f_{r<s}(\sigma(1))} 
	\cdots 
	\xi_{\alpha_{n - 2},\,f_{r<s}(\sigma(n - 2))} 
	\right) 
\nonumber \\ 
	&& \quad 
	\times 
	\left( 
	\sum_{\sigma \in {\mathfrak S}_{n - 2}} 
	\varepsilon(\sigma)\,  
	\xi_{\alpha_1,\,f_{r^\prime < s^\prime}(\sigma(1))} 
	\cdots 
	\xi_{\alpha_{n - 2},\,f_{r^\prime < s^\prime}(\sigma(n - 2))} 
	\right) \, . 
\nonumber \\ 
\end{eqnarray} 
By inserting this expression into Eq.~(\ref{eq:B4_ito_det}), 
$B_{\alpha,\,\beta;\,\alpha^\prime,\,\beta^\prime}$ 
is expressed in a simple form 
\begin{equation} 
	B_{\alpha,\,\beta;\,\alpha^\prime,\,\beta^\prime} 
	= 
	\sum_{\alpha_1 < \cdots < \alpha_{n - 2}} 
	w_{\alpha_1} \cdots w_{\alpha_{n - 2}}\, 
	\Theta(\alpha,\,\alpha^\prime;\,\alpha_1,\,\cdots,\,\alpha_{n - 2})\, 
	\Theta(\beta,\,\beta^\prime;\,\alpha_1,\,\cdots,\,\alpha_{n - 2}) \, , 
\end{equation} 
with 
\begin{eqnarray} 
	\Theta(\alpha,\,\beta;\,\alpha_1,\,\cdots,\,\alpha_{n - 2})\, 
	&\equiv& 
	\sum_{1 \le r < s \le n} 
	(-1)^{r + s + 1} 
	\left( 
	\xi_{\alpha,\,r} \xi_{\beta,\,s} 
	- \xi_{\alpha,\,s} \xi_{\beta,\,r} 
	\right) 
\nonumber \\ 
	&& \quad 
	\times 
	\sum_{\sigma \in {\mathfrak S}_{n - 2}} 
	\varepsilon(\sigma)\, 
	\xi_{\alpha_1,\,f_{r<s}(\sigma(1))} \cdots 
	\xi_{\alpha_{n - 2},\,f_{r<s}(\sigma(n - 2))} \, . 
\nonumber \\ 
\end{eqnarray} 
The next task is to examine if  
$\Theta(\alpha,\,\beta;\,\alpha_1,\,\cdots,\,\alpha_{n - 2})$ 
coincides with 
$A(\alpha,\,\beta,\alpha_1,\,\cdots,\,\alpha_{n - 2})$. 
As was done to derive the first equality of Eq.~(\ref{eq:B_equalTo_A}), 
we replace the sum over the permutations of degree $(n - 2)$ 
on the indices of loops 
with that over the permutations on the chain indices to get 
\begin{eqnarray} 
 \Theta(\alpha,\,\beta;\,\alpha_1,\,\cdots,\,\alpha_{n - 2}) 
 &=& 
 \sum_{1 \le r < s \le n} 
  (-1)^{r + s + 1} 
  \left( 
     \xi_{\alpha,\,r} \xi_{\beta,\,s} 
   - \xi_{\alpha,\,s} \xi_{\beta,\,r} 
  \right) 
  \nonumber \\ 
 && \times 
 \sum_{\sigma \in {\mathfrak S}_{n - 2}} 
  \varepsilon(\sigma)\, 
  \xi_{\alpha_{\sigma(1)},\,f_{r<s}(1)} 
  \cdots 
  \xi_{\alpha_{\sigma(n - 2)},\,f_{r<s}(n - 2)} 
   \nonumber \\ 
 &=& 
 \sum_{1 \le r < s \le n} 
  \sum_{\sigma \in {\mathfrak S}_{n - 2}}  
  (-1)^{r + s + 1} \varepsilon(\sigma) 
   \nonumber \\ 
 && \qquad 
  \times 
  \xi_{\alpha_{\sigma(1)},\,1} 
  \cdots 
  \xi_{\alpha_{\sigma(r - 1)},\,r - 1}\,  
  \xi_{\alpha,\,r}\, 
  \xi_{\alpha_{\sigma(r)},\,r + 1} \cdots 
  \nonumber \\ 
 && \qquad 
  \times 
  \xi_{\alpha_{\sigma(s - 2)},\,s - 1}\,  
  \xi_{\beta,\,s}\,  
  \xi_{\alpha_{\sigma(s - 1)},\,s + 1} 
  \cdots 
  \xi_{\alpha_{\sigma(n - 2)},\,n} 
   \nonumber \\ 
 && \ 
 +  
 \sum_{1 \le r < s \le n} 
  \sum_{\sigma \in {\mathfrak S}_{n - 2}}  
  (-1)^{r + s + 2} \varepsilon(\sigma) 
   \nonumber \\ 
 && \qquad 
  \times 
  \xi_{\alpha_{\sigma(1)},\,1} 
  \cdots 
  \xi_{\alpha_{\sigma(r - 1)},\,r - 1}\,  
  \xi_{\beta,\,r}\,  
  \xi_{\alpha_{\sigma(r)},\,r + 1} \cdots 
  \nonumber \\ 
 && \qquad 
  \times 
  \xi_{\alpha_{\sigma(s - 2)},\,s - 1}\,  
  \xi_{\alpha,\,s}\,  
  \xi_{\alpha_{\sigma(s - 1)},\,s + 1} 
  \cdots 
  \xi_{\alpha_{\sigma(n - 2)},\,n} 
   \nonumber \\ 
 &=& 
 \sum_{\sigma \in {\mathfrak S}_{n}} 
  \varepsilon(\sigma)\, 
  \xi_{\alpha,\,\sigma(1)}\,\xi_{\beta,\,\sigma(2)}\, 
  \xi_{\alpha_1,\,\sigma(3)} \cdots 
  \xi_{\alpha_{n - 2},\,\sigma(n)} 
  \nonumber \\ 
 &=& 
 A(\alpha,\,\beta,\,\alpha_1,\,\cdots,\,\alpha_{n - 2}) 
  \, . 
\end{eqnarray} 
 The third equality is given by using the fact that 
the sequences of permutations 
\begin{eqnarray} 
 && 
 \left[ 
  \alpha_{\sigma(1)},\,\cdots,\,\alpha_{\sigma(r - 1)},\, 
  \alpha,\,\alpha_{\sigma(r)},\,\cdots,\, 
  \alpha_{\sigma(s - 2)},\,\beta,\,\alpha_{\sigma(s - 1)},\,
  \cdots,\,\alpha_{\sigma(n - 2)} 
 \right] 
  \nonumber \\ 
 && \quad 
 \mapsto 
 \left[ 
  \alpha,\,\beta,\, 
  \alpha_{\sigma(1)},\,\cdots,\,\alpha_{\sigma(r - 1)},\, 
  \alpha_{\sigma(r)},\,\cdots,\,\alpha_{\sigma(s - 2)},\, 
  \alpha_{\sigma(s - 1)},\,\cdots,\,\alpha_{\sigma(n - 2)} 
 \right] 
  \nonumber \\ 
 && \quad 
 \mapsto 
 \left[ 
  \alpha,\,\beta,\,\alpha_1,\,\cdots,\,\alpha_{n - 2} 
 \right] \, , 
  \nonumber \\ 
 && 
 \left[ 
  \alpha_{\sigma(1)},\,\cdots,\,\alpha_{\sigma(r - 1)},\, 
  \beta,\,\alpha_{\sigma(r)},\,\cdots,\, 
  \alpha_{\sigma(s - 2)},\,\alpha,\,\alpha_{\sigma(s - 1)},\,
  \cdots,\,\alpha_{\sigma(n - 2)} 
 \right] 
  \nonumber \\ 
 && \quad 
 \mapsto 
 \left[ 
  \alpha,\,\beta,\, 
  \alpha_{\sigma(1)},\,\cdots,\,\alpha_{\sigma(r - 1)},\, 
  \alpha_{\sigma(r)},\,\cdots,\,\alpha_{\sigma(s - 2)},\, 
  \alpha_{\sigma(s - 1)},\,\cdots,\,\alpha_{\sigma(n - 2)} 
 \right] 
  \nonumber \\ 
 && \quad 
 \mapsto 
 \left[ 
  \alpha,\,\beta,\,\alpha_1,\,\cdots,\,\alpha_{n - 2} 
 \right] \, , 
\end{eqnarray} 
where the two steps in the first case 
produce signatures $(-1)^{r + s + 1}$ and $\varepsilon(\sigma)$ 
respectively, 
and the two steps in the second case 
produce $(-1)^{r + s + 2}$ and $\varepsilon(\sigma)$ respectively. 
 They combine to form all possible permutations of degree $n$. 
 In this way, we finally reach at a compact formula 
for $B_{\alpha,\,\beta;\,\alpha^\prime,\,\beta^\prime}$; 
\begin{eqnarray} 
 B_{\alpha,\,\beta;\,\alpha^\prime,\,\beta^\prime} 
 &=& 
 \sum_{\alpha_1 < \cdots < \alpha_{n - 2}} 
  z_{\alpha_1} \cdots z_{\alpha_{n - 2}} 
  \nonumber \\ 
 && \qquad \qquad 
 \times 
 A(\alpha,\,\alpha^\prime,\,\alpha_1,\,\cdots,\,\alpha_{n - 2})\,  
 A(\beta,\,\beta^\prime,\,\alpha_1,\,\cdots,\,\alpha_{n - 2}) \, . 
  \label{eq:new_formula_for_Babcd}
\end{eqnarray} 
 This formula provides a way to calculate 
$B_{\alpha,\,\beta;\,\alpha^\prime,\,\beta^\prime}$ 
algebraically. 

 The cases (I), (II\,a), (II\,b), (II\,c)\, (II\,d) 
contain $B_{\alpha,\,\beta;\,\alpha^\prime,\,\beta^\prime}$, 
where at least two of the chain indices coincide with each other. 
 Actually such polynomials can be calculated 
much rapidly instead of using Eq.~(\ref{eq:new_formula_for_Babcd}). 
 From Eq.~(\ref{eq:Baa}) for $B_{\alpha\,\alpha}$  
and the definition (\ref{eq:BbyChain}) 
of $B_{\alpha\,\beta}$, 
$B_{\alpha,\,\alpha;\,\alpha^\prime,\,\beta^\prime}$ 
can be written as 
\begin{eqnarray} 
 B_{\alpha,\,\alpha;\,\alpha^\prime,\,\beta^\prime} 
 &=& 
 \frac{1}{U} 
 \left[ 
  \frac{\del U}{\del w_\alpha}\, 
  \left( 
   \sum_{r,\,s = 1}^{n} 
    \xi_{\alpha^\prime,\,r}\,U (U^{-1})_{r s}\, 
    \xi_{\beta^\prime,\,s} 
  \right) 
 \right. 
  \nonumber \\ 
 && \qquad 
 \left. 
  - 
  \sum_{r,\,s,\,u,\,v = 1}^{n} 
   \xi_{\alpha^\prime,\,r}\, 
   U (U^{-1})_{r u}\,\xi_{\alpha,\,u}\,\xi_{\alpha,\,v} 
   U (U^{-1})_{v s}\,\xi_{\beta^\prime,\,s} 
 \right] 
  \nonumber \\ 
 &=& 
 \frac{\del U}{\del w_\alpha}\, 
 \left( 
  \sum_{r,\,s = 1}^{n} 
   \xi_{\alpha^\prime,\,r}\,(U^{-1})_{r s}\, 
   \xi_{\beta^\prime,\,s} 
 \right) 
  \nonumber \\ 
 && \qquad 
 - U 
 \sum_{r,\,s,\,u,\,v = 1}^{n} 
  \xi_{\alpha^\prime,\,r}\, 
  (U^{-1})_{r u}\,\frac{\del U_{uv}}{\del w_\alpha} 
  (U^{-1})_{v s}\,\xi_{\beta^\prime,\,s} 
  \nonumber \\ 
 &=& 
 \sum_{r,\,s = 1}^{n} 
  \xi_{\alpha^\prime,\,r}\, 
  \frac{\del \left(U (U^{-1})_{r s}\right)}{\del w_\alpha}\, 
  \xi_{\beta^\prime,\,s} \, . 
\end{eqnarray}  
 This yields an equality  
\begin{equation} 
 B_{\alpha,\,\alpha;\,\alpha^\prime,\,\beta^\prime} 
 = 
 \frac{\del B_{\alpha^\prime\,\beta^\prime}}{\del w_\alpha} \, . 
  \label{eq:formulaForBaabc}
\end{equation} 
 This equality also follows from 
the expression (\ref{eq:new_formular_for_Bfunction}) 
for $B_{\alpha^\prime\,\beta^\prime}$ 
and Eq.~(\ref{eq:new_formula_for_Babcd}). 
 While Eq.~(\ref{eq:new_formula_for_Babcd}) requires 
the calculation of 
two $A(\alpha,\,\alpha^\prime,\,\alpha_1,\,\cdots,\,\alpha_{n - 2})$, 
the formula (\ref{eq:formulaForBaabc}) allows us to obtain 
$B_{\alpha,\,\alpha;\,\alpha^\prime,\,\beta^\prime}$ 
simply by looking for the monomials containing $w_\alpha$ 
in $B_{\alpha^\prime,\,\beta^\prime}$. 
 The use of Eq.~(\ref{eq:formulaForBaabc}) 
is very efficient for the calculation of 
$B_{\alpha,\,\alpha;\,\alpha^\prime,\,\beta^\prime}$. 
Finally, these results can be translated into those for 
$C_{ij}$ through Eqs.~(\ref{eq:defCPoly}), 
(\ref{eq:cij:defc}), and (\ref{eq:DefOfB4Tensor}).
%----------------------------------------------------------------
%----------------------------------------------------------------

%================================================================

%================================================================
% REFERENCES
%================================================================

%================================================================

%----------------------------------------------------------------
%----------------------------------------------------------------
%----------------------------------------------------------------
%----------------------------------------------------------------
\end{document}